\documentstyle[preprint,prb,aps,overcite,tighten]{revtex}

\begin{document}

\title{Multi-Component Quantum Hall Systems: The
Sum of Their Parts and More}

\author{S.M. Girvin and A.H. MacDonald}

\address{Department of Physics, Indiana University,
 Bloomington, Indiana 47405 USA}


\maketitle

{\tightenlines
\begin{abstract}

The physics of the fractional quantum Hall effect is the
physics of interacting electrons confined to a macroscopically
degenerate Landau level.  In this Chapter we discuss the
theory of the quantum Hall effect in systems where the electrons
have degrees of freedom in addition to the two-dimensional orbital
degree of freedom.  We will be primarily interested in the
situation where a finite number of states, most-commonly
two, are available for each orbital state within a degenerate
Landau level and will refer to these systems as multi-component systems.
Physical realizations of the additional degree of freedom include the
electron spin, the valley index in multi-valley semiconductors,
and the layer index in multiple-quantum-well systems.
The consideration of multi-component systems expands the
taxonomy of incompressible states and fractionally charged
excitations, and for example, leads to the appearance of
fractions with even denominators.  More interestingly,
it also leads us to new physics, including novel spontaneously
broken symmetries and in some cases, finite temperature phase transitions.
We present an introduction to this rich subject.

\end{abstract}
}
\vfill
\noindent
To appear in: {\em Novel Quantum Liquids in Low-Dimensional Semiconductor
Structures}, edited by Sankar Das Sarma and Aron Pinczuk (Wiley, New York,
1995).
\newpage
\pagenumbering{roman}
{\tighten
\tableofcontents
}
\newpage
\pagenumbering{arabic}

\section{Introduction}
\label{sec:intro}
The fractional quantum Hall
effect\cite{smgbook,ahmbook,chakrabortybook,stonebook}
is a remarkable example of strong
correlations in a two-dimensional electron gas (2DEG).  
In zero magnetic field,
a dimensionless measure of the strength of the Coulomb interaction
for a system with dielectric constant $\epsilon$,
Fermi energy $\epsilon_{\rm F}$ and Fermi wavevector $k_F$ is 
\begin{equation}
\lambda = \frac{e^2 k_{\rm F}/\epsilon}{\epsilon_{\rm F}},
\end{equation}
which is small in the limit of high density.  In this limit one can
frequently treat the effects of the Coulomb interaction perturbatively.
Physically this can be visualized as being due to the electrons moving
rapidly past each other at the Fermi velocity and thus not scattering
so strongly as they would at lower densities.
A strong magnetic field completely changes this situation.
Semiclassically, the rapid motion of the electrons is
converted into circular cyclotron orbits.  The particles now scatter
strongly from each other and in fact semiclassically do not move except under
the ${\bf E}\times{\bf B}$ drift induced by their mutual interactions.
A full quantum treatment of the
motion shows that the kinetic energy is quenched and now occurs only
in discrete values $(n+1/2)\hbar\omega_{\rm c}$, where $n$ is the
Landau level index.  Since the kinetic energy within a given Landau level
is completely degenerate, the Coulomb interaction inevitably induces
highly non-perturbative effects.  (The Landau level degeneracy
is $N_{\phi} = B A / \Phi_0$ where $\Phi_0 = h c /e $ is
the magnetic flux quantum, and $A$ is the area of the system.)

The essential feature of the fractional quantum Hall effect is a
condensation of the electrons into  special highly correlated
states\cite{RBLchap7}
which minimize the Coulomb energy by having the electrons avoid each
other as much as possible.  These states are characterized by an unusual
topological order\cite{SMGchap10,ODLRO,WEN_topological,moore+read}
which costs a
finite amount of energy to break.  Hence the fluid is incompressible and
has an excitation gap both for its charged\cite{RBLchap7} and
neutral\cite{GMP,SMGchap9} excitations.  P.~W.~Anderson has characterized
this state as a Mott insulator induced by the magnetic
field.\cite{PWA_privcomm}

The essence of this phenomenon is captured
in a remarkable class of wave functions first constructed by
R.~B.~Laughlin
\begin{equation}
\psi_m(Z_1,Z_2,...,Z_N) = \prod_{i<j}^N \left(Z_i - Z_j\right)^m
\exp{\left\{-\frac{1}{2}\sum_k^N |Z_k|^2\right\}}.
\label{RBLpsi}
\end{equation}
Here, since we are in two dimensions, we are using the dimensionless
complex number
$Z=(x+iy)/\ell$ to represent the position vector $(x,y)$
in units of the magnetic length $\ell = \sqrt{\hbar c/eB}$.
This wave function describes spinless fermions in the lowest Landau level
(in the symmetric gauge).  To satisfy the analyticity requirement placed
on the wave function by the constraint of being in the lowest Landau
level,\cite{girvin-jach}
the parameter $m$ must be an integer.  To satisfy the antisymmetry
requirement for fermions, $m$ must be odd.
Laughlin's plasma analogy\cite{RBLchap7} shows that the
parameter $m$ fixes the Landau level filling factor to be $\nu \equiv
N/N_{\phi} = 1/m$.
Experiment\cite{experiment}
has indeed observed gapped quantum Hall states at filling factors
$\nu=1, 1/3$, and $1/5$.

It is clear that Laughlin's wave function builds in good correlations
because it vanishes as $|{\bf r}_i - {\bf r}_j|^m$ when any two particles $i$
and $j$ approach each other.  Thus there is only a small amplitude for
the particles to be near each other, and the Coulomb energy is lowered.
Note that no pair of particles ever has relative angular momentum less than
$m$.  Hence the Laughlin function is a zero-energy
exact eigenfunction for
Hamiltonians with the appropriate
finite number of non-zero Haldane pseudopotentials\cite{FDMHchap8}, $V_n$:
\begin{equation}
H = \sum_{n<m} V_n \sum_{i<j} P_n[i,j],
\end{equation}
where $P_n[i,j]$ is the lowest Landau level (LLL)
projection operator for the relative angular
momentum state $n$ of particles $i$ and $j$.  In the lowest Landau level,
relative angular momentum is proportional to the square of the
separation.\cite{girvin-jach}
Hence the Laughlin wave function is very nearly an exact ground state for
any sufficiently short-range repulsive interaction.

In addition to the primary filling fractions
$\nu=1/m$, numerous other fractions have
been observed, all of which (for single component systems)
have odd denominators (again because of the Pauli principle).  
These phases have been explained in terms of a hierarchical picture using
bosonic\cite{FDMHchap8}, anyonic\cite{bert_anyonic}, and
fermionic\cite{RBLchap7,SMG_hier} representations.  More recently
Jain\cite{jain_CF,jain_thisbook} has discussed an appealing
 composite fermion picture.  N. Read has argued that
all of these representations are mathematically 
equivalent\cite{readhier} and contain the same physics.
Which representation is most convenient depends on circumstances.
Jain's approach has inspired several new and important
experiments.\cite{jain_thisbook}

Our purpose here is to consider the nature of the various phases
which can occur in multi-component systems.  There are several physically
different realizations of systems with extra degrees of freedom which
require a multi-component representation,  and
important early work on this problem was done by 
Halperin.\cite{halperinz1,macdsurface}

The first and simplest example is that
of ordinary electron spin.  In free space for electrons with $g$ factor 2
(i.e., neglecting QED corrections) the Zeeman splitting $g\mu_{\rm B}B$ is
precisely equal to the Landau level splitting $\hbar\omega_{\rm c}$.  If
this were true in quantum Hall samples, then even a
non-interacting system with filling factor
$\nu=1$ would have a large excitation gap for flipping spins and the ground
state would be fully polarized at low temperatures.
In this case spin excitations are frozen out and we can treat the electrons
as being effectively spinless.  However in the solid state environment of
the 2DEG, two factors conspire to to make the effective g-factor much
smaller in many semiconductors and particularly in GaAs samples in
which almost all fractional quantum Hall studies have been done.
The first is that the small
effective mass ($m^* \sim 0.068$ in GaAs) increases the cyclotron energy by
a factor of approximately 15.  Secondly, spin-orbit coupling causes the
spins to tumble and reduces their coupling to the external magnetic
field by roughly a factor of 4.
Thus the ratio of the Zeeman splitting to cyclotron splitting
is reduced from unity to about 0.02
in GaAs.  The spin-orbit contribution is pressure dependent and may
also
be further reduced as a result of size quantization effects in narrow
quantum wells.\cite{pressure-tuned}

For small enough $g$ factor and weak enough magnetic fields, spin fluctuations
become an important dynamical degree of freedom and we must use a
two-component wave function to describe the system.  While the Coulomb
force is spin-independent, we shall see below that
exchange effects can lead to spontaneous
ferromagnetism as well as to gapped `local singlet' spin liquids, which
are, loosely speaking, itinerant antiferromagnets. [Unlike antiferromagnets
however, they do not break translation symmetry nor do they have gapless
Goldstone modes).  Ferromagnetism turns out to be important at filling
factor $\nu=1$ where Coulomb exchange effects are much more important
than originally realized.\cite{sondhi,fertig,barrett_prl,barrett_science}

A different class of quantum Hall states with gaps has been observed
experimentally\cite{clark86_5/2,willet87_5/2,gammel88_5/2,jpe_tilt_5/2}
 at filling fractions with even denominators such as $\nu=5/2$.  These
have been 
argued\cite{haldane-rezayi,yosh89,five-halves-theory,moore+read,greiter}  
to be either manifestations of spin liquid states or 
special `p-wave pairing' spin-polarized states known as pfaffian states.
One useful signature of spin effects in 2DEG's
is that they are sensitive to tilts of the magnetic field.  To a first
approximation, the orbital degrees of freedom are sensitive only to the
perpendicular component of the field  because of the 2D confinement,
while the Zeeman splitting is proportional to the total magnetic field.
(However, the coupling of orbital degrees of freedom to the parallel component
of the magnetic field is not
completely negligible in typical quantum wells\cite{nickila} and this
can considerably complicate the situation.)

A second example of a multi-component system
is found in silicon where the conduction band minimum occurs
not at the $\Gamma$ point (the zone center) but rather at six symmetry
equivalent points lying near the zone boundary along the
principal cubic directions.  Thus electrons doped into the conduction band
of Si must be described by a six-component wave function (if we ignore spin).
The presence of the oxide barrier in a Si MOSFET device and the enormous
electric field perpendicular to it (which is used to confine the electrons
into the inversion layer) breaks the cubic symmetry and lowers the energy
of two of the valleys. For typical electron densities, only these two
valleys are occupied, thus yielding a system
which is effectively two-component and has SU(2) symmetry just like a
spin-1/2 system.\cite{valley-degen}

A third example which will be extensively discussed here and in the Chapter
in this volume
by J.~P.~Eisenstein\cite{JPEchap} occurs in double quantum well
structures.\cite{murphyetal}
With modern MBE techniques it is possible to grow GaAs heterostructures
containing two 2DEG's separated by a distance comparable to the spacing
between electrons within each layer.  Remarkably, it is also possible to make
separate electrical contact to each layer.  A closely related system is
a single wide quantum well in which the two lowest electric subbands are
nearly degenerate.\cite{mansour}
We will make the (only approximately correct) assumption throughout our
discussion that the low energy physics of a single wide well can be mapped
onto that of a double well with appropriately chosen parameters.
In all two component systems it is
useful to define a pseudospin representation in which spin up and down
refer to the two possible values of the layer index (or subband index)
for each electron.\cite{macd-expt}  We will frequently frame our
discussion in a spin or pseudospin language; the reader should be aware that
the discussion applies equally well to double-layer systems.  When
the distinction between the U(1) and SU(2) symmetries of the
interaction term in the Hamiltonian is important, we will say so.

Because the intra- and inter-layer Coulomb
matrix elements are different, double well systems do not have full SU(2)
symmetry, but rather only U(1) symmetry associated with the conservation of
the charge difference between the two layers (assuming there is no
interlayer
tunneling).\cite{macd-expt,Wen-Zee_double,ezawa,ezawa_95,usPRL,uslongI,uslongII}
These systems exhibit gapped quantum Hall states with both even and odd
denominators.  In addition, there are gapless XY ordered phases.  These
phases are destroyed above some critical temperature by a
Kosterlitz-Thouless phase transition.
This phase transition (not yet observed experimentally) is the first
example of a finite temperature phase transition in a quantum Hall system.

This Chapter is organized as follows.  In Section~\ref{sec:multi-psi}
we introduce the physics of incompressible states in multi-component
cases by presenting the appropriate generalizations of Laughlin's
many-particle wavefunctions.  Section~\ref{sec:chern-simons} briefly
directs the reader to references on the Chern Simons effective field theory
approach to these problems.
Section~\ref{sec:frac-charge}
discusses fractional charges  in multicomponent systems.
Section~\ref{sec:SMA} discusses
collective modes from the point of view of the single-mode approximation.
In both these sections we will see that some of the multi-component
wavefunctions have correlation functions whose qualitative properties
deviate from the norm.  These differences lead to changes in the nature
of the fractionally charged excitations and in the collective modes.
The qualitative difference is associated with a broken symmetry which
occurs in some cases as we discuss in Section~\ref{sec:broken-symm}.
In our view, it is in the properties of these broken symmetry states
that some of the most interesting new physics in multi-component
quantum Hall states is revealed.  Section~\ref{sec:field-theory}
discusses the field-theoretic gradient expansion approach to the broken
symmetry case.
In double-layer systems, the broken symmetry state
has spontaneous phase coherence between the electrons in different
layers even when these layers are isolated apart from
inter-layer Coulomb interactions.
Section~\ref{sec:tunneling} deals with the external symmetry breaking
introduced by interlayer tunneling.
Spontaneous coherence in double-layer systems leads to remarkable
effects upon tilting the magnetic field away from the normal
in a double layer system.  Some of these effects are
discussed in Section~\ref{sec:parallelB}.
Finally Section~\ref{sec:summary} presents a summary of the central ideas.

\section{Multi-Component Wave Functions}
\label{sec:multi-psi}

In this section we discuss wavefunctions for spin or pseudospin $1/2$
particles which can be written in the symmetric gauge in the form
\begin{equation}
\Psi [Z:\chi ]=A[\Phi [Z] \alpha_1 \cdots \alpha_{N_\uparrow} \beta_{[1]}
\cdots \beta_{[N_\downarrow]}],
\end{equation}
where $A$ is the antisymmetrization operator, $[i]\equiv
N_\uparrow +i$, and $\alpha_k$ and $\beta_k$ are the spinors for the $k$th
electron aligned respectively parallel and antiparallel to the Zeeman
field (which we take to be in the ${\bf\hat z}$ direction).
Thus $\Phi [Z]$ is the
orbital wave function for the spin configuration in which the first
$N_\uparrow$ electrons have spin ``up" and the remaining electrons have
spin ``down".\cite{realize:up-down}

The two-component orbital wave functions originally proposed by
Halperin\cite{halperinz1,macdsurface} have a form analogous to the
Laughlin functions
\begin{eqnarray}
 && \Phi_{m,m',n} [Z]=\prod_{i<j\leq N_\uparrow} (Z_i-Z_j)^m
\prod_{k<l\leq N_\downarrow} (Z_{[k]}-Z_{[l]})^{m'}\nonumber\\
 && \phantom{\Phi}\times \prod_{a=1}^{N_\uparrow}
\prod_{b=1}^{N_\downarrow} (Z_a-Z_{[b]})^n \prod_{s=1}^N
\exp{\left\{-|Z_s|^2/4\right\}},
\label{eq:halperin-multi-psi}
\end{eqnarray}
where $Z_k=(x_k+iy_k)/\ell$ is the 2D layer coordinate of the $k$th electron
expressed as a complex number, and $m$ and $m'$ are odd integers.
$\Phi_{m,m',n} [Z]$ excludes relative angular momenta less than $m$
between up-spins, less than $m'$ between
down-spins, and less than $n$ between an up-spin
and a down-spin.  Hence the same arguments presented above for the Laughlin
wave function can be used to motivate the idea that the Halperin wave functions
are good approximations to the ground state for short-range repulsive
potentials.

In the single component case, Laughlin wavefunctions are
expected to accurately approximate the ground state for $m =3$ and
$m=5$ ($\nu = 1/3$ and $\nu =1/5$); at smaller filling factors the Wigner
crystal state takes over.  In the two-component case, the realization
of a Laughlin state at a particular filling factor may be dependent on
other parameters of the system such as the layer separation in
double-layer systems.
One slight complication is presented by the SU(2) symmetric
case.\cite{FDMHchap8,yosh88}
  Here
the total spin $S_{\rm T}$ commutes with the Hamiltonian and one may, without
loss of generality, require that the energy eigenstates
simultaneously be eigenstates of
$S_{\rm T}^2$.  This requirement is not satisfied in general by the
Halperin wave functions.  It is easy to pick out two special cases (among
others) that do work however.
The choice $m=m'=n$ yields a fully
antisymmetric spatial wave function and hence implies a fully symmetric
spin function.  This immediately tells us that we have a fully aligned
ferromagnetic state with total spin quantum number
$S = N/2$.  On the other hand,
the choice $\{m,m',n\} = \{1,1,0\}$
corresponds to a simple Slater determinant with both spin states
of the lowest Landau level fully  occupied, giving $\nu=2$.
Hence it is automatically a spin singlet.  We can generalize this
to $\{m,m',n\}=\{m,m,(m-1)\}$ since this corresponds to simply multiplying
the filled Landau level function by a fully symmetric spatial polynomial.
Thus these states are also spin singlets.

Using the fact that every extended single-particle orbital
in the lowest Landau level involves a polynomial with a fixed (average)
density of zeros\cite{halperinz1,macdsurface}
 given by $B/\Phi_0$ (where $\Phi_0\equiv hc/e$ is the flux quantum)
we may derive a pair of equations for the density
$\rho$ of each component
\begin{eqnarray}
\frac{B}{\Phi_0} &=& m\rho_\uparrow + n\rho_\downarrow,\nonumber\\
\frac{B}{\Phi_0} &=& m'\rho_\downarrow + n\rho_\uparrow.
\end{eqnarray}
\null From this we obtain for the filling factors
\begin{eqnarray}
\nu_\uparrow &=& \frac{m' - n}{mm' - n^2},\nonumber\\
\nu_\downarrow &=& \frac{m - n}{mm' - n^2}.
\label{eq:partial_fillings}
\end{eqnarray}
Partial and total filling factors for some of these two-component
Laughlin states are listed in Table~\ref{table1}.  At these partial
filling factors $\Phi_{m,m',n} [Z]$ is unique in the sense that
it is the {\em only} wave function which
excludes its corresponding low relative angular momentum channels and,
just as in the case of Laughlin states in one-component systems, we may
expect that these wave functions will be nearly exact groundstates for
any sufficiently short-ranged repulsive
interaction.
In the SU(2) invariant case it seems that we should\cite{tosatti} however
require
that the wave functions be eigenstates of the total spin operator
$S_{\rm T}$
and that their Zeeman energy be not too unfavorable.  There is numerical
evidence for instance,
that the $\{3,3,2\}$
state which has $\nu\equiv \nu_\uparrow + \nu_\downarrow =2/5$
has a lower energy
than the usual hierarchical state if (and for typical field strengths,
only if) one ignores the Zeeman
energy.  For further references and detailed discussion on this point
and other topics related to spin in the FQHE, the reader is
directed to the book by Chakraborty and Pietil\"ainen.\cite{chakrabortybook}

We note that Eq.(\ref{eq:partial_fillings}) is ill-defined if
$mm'-n^2$ vanishes as it does, for example, in
the fully ferromagnetic $\{m,m,m\}$ states.
In this case, however, one can compute the filling factor by simply noting
that the fully spin aligned system has an orbital wave function equivalent
to the Laughlin function at total filling factor $\nu=1/m$.  The relative
filling factors of the two components is necessarily ill-defined because of
the SU(2) rotational symmetry.  There are $2S+1 = N+1$ orthogonal but
macroscopically
degenerate states differing only by their $S^z$ quantum number.  Thus we
have
\begin{eqnarray}
\nu_\uparrow + \nu_\downarrow &=& \frac{1}{m},\nonumber\\
\nu_\uparrow - \nu_\downarrow &=& \frac{2S^z}{Nm}.
\end{eqnarray}
The degeneracy of these states leads to a broken symmetry which we discuss
in greater detail in Section~\ref{sec:broken-symm}.

Generalizations of the Halperin states can be made to the case of an
arbitrary number of components.\cite{macd_science,qiu-joynt-macd,hanna-macd}
This would have application for example
to a superlattice of closely
spaced quantum wells, should these become technologically feasible
to produce at some point in the future.
It is known
experimentally\cite{clark86_5/2,willet87_5/2,gammel88_5/2,jpe_tilt_5/2}
 that there exists an incompressible (but unusually delicate) Hall
state at filling factor $\nu=5/2$.  This state has also been observed
numerically for various artificially chosen
 interaction models, but not however, for
a pure Coulomb 
interaction.\cite{yosh89,five-halves-theory,belkhir_thesis,chakrabortybook}
It has been argued that the ground state
at this filling factor is not fully spin polarized because the Hall
state is easily destroyed by tilting the magnetic field at constant filling
fraction.\cite{jpe_tilt_5/2} 
In principle, the same state should be observed at filling
$\nu=1/2$ because $\nu=5/2 = 2 + 1/2$
has both spin states of the LLL filled and these electrons are essentially
inert, leaving an effective filling factor
$\nu_{\rm eff}=1/2$ in the next Landau level.
However the analogous state is not observed at $\nu=1/2$
because, to reach this
lower filling factor (in the same GaAs sample), it is necessary to increase
the perpendicular component of the magnetic field by a factor of 5.  This
increases the Zeeman energy advantage of the fully spin polarized state
(which is gapless and does not exhibit a Hall plateau)
and makes it the ground state.  Thus non-observation
of a quantum Hall state at $\nu=1/2$ provides additional evidence that the
$\nu=5/2$ state is not fully spin polarized.  As described in
the Chapter in this volume by Eisenstein, tilted field experiments
can also help identify spin-reversed quasi-particle excitations above
polarized ground states.\cite{jpe_reverse_qp,chakrabortybook}

Haldane and Rezayi\cite{haldane-rezayi} have proposed a
two-component spin-singlet wave function
to explain the existence of the $\nu=5/2$ quantum Hall plateau.  This
wave function is an exact ground state for the so-called
`hollow-core' model\cite{haldane-rezayi}
 and may be written in two different ways by modifying two different
 Halperin states, each of which has filling factor one
 half.\cite{foot:simplicity1}
The first uses the Halperin fermionic $\{3,3,1\}$ function
\begin{equation}
\Phi_{\rm HC} = \Phi_{331}\,\, {\rm per}|M|,
\end{equation}
where $M$ is an $N/2\times N/2$ matrix whose $ij$ element is given by
\begin{equation}
M_{ij} = \left(Z_i - Z_{[j]}\right)^{-1}.
\end{equation}
The permanent of the matrix, ${\rm per}|M|$, is by definition,
just like the determinant
except that there are no minus signs for odd permutations.
The subtle effect of this permanent on the
wave function is to cause it to be an eigenfunction of total spin
(with $S=0$) {\em without} changing the density (at least
in the thermodynamic limit).

Note that the permanent causes some up and down spin particles to have
a finite probability of having relative angular momentum zero.  It turns
out that no particles  ever have relative angular momentum 1 in this
state.  Hence
this wave function is an exact zero energy singlet ground state for
the `hollow core' potential model
\begin{equation}
V_m = V \delta_{m,1}.
\end{equation}
Despite the unphysical appearance of this model, it has been argued that it
might capture the correct physics when the form of the effective
pseudopotentials in the second Landau level 
is taken into account.\cite{haldane-rezayi}

A second way to write the same state uses the (222) bosonic Halperin
wave function
\begin{equation}
\Phi_{\rm HC} = \Phi_{222} {\rm Det}|{\tilde M}|,
\end{equation}
where $M$ is an $N/2\times N/2$ matrix whose $ij$ element is given by
\begin{equation}
M_{ij} = \left(Z_i - Z_{[j]}\right)^{-2}.
\end{equation}
A curious mathematical identity\cite{haldane-rezayi} allows one to show
that these two representations are precisely equivalent.  

It is possible that orbital effects\cite{nickila} confuse the 
tilted field test for spin unpolarized states and the $\nu=5/2$ is actually
spin-polarized.  A competing spin-polarized `pfaffian' state developed by 
Read\cite{moore+read} and also studied by Greiter et al.\cite{greiter}
is a kind of `p-wave' paired state.
It is not known for certain at this point what the true nature of the very
delicate 5/2 state is.\cite{belkhir_thesis,JPEchap}  It may be one of the 
proposed states\cite{yosh89,five-halves-theory,moore+read,haldane-rezayi} 
or something completely unknown.

Jason Ho has recently considered interesting
connections between the internal order in
these types of wave functions and analogous order in superfluid
$^3$He.\cite{jasonho-deform}  In particular he has discussed the
"incompressible" deformation of the various states into each other,
connecting for example the $\{3,3,1\}$ state and the pfaffian 
state.\cite{jasonho-deform,greiter,Halperin_Newport}

\section{Chern Simons Effective Field Theory}
\label{sec:chern-simons}

One interesting approach to the quantum Hall effect in general, and
multi-component systems in particular, is the Chern Simons effective
field theory.  Unfortunately space  limitations prevent us from discussing
this approach in any detail.
The reader is directed to the Chapter in this volume
by B. I. Halperin,\cite{halperinchap} the references therein
 and to the many references in what is
now a vast literature.%
\cite{kaneandlee,sondhi,renndbl,du,zhang_cs,lz,zhang-macd-collective,%
lopez,uslongI,ezawa,ezawa_95,Wen-Zee_double,birman,bonesteel,jasonho,Dziarmaga,%
zee_review,froelich_review,lee-wen}
A succinct and introductory summary of the bosonic representation
for double layer systems is given in
Ref.[\citen{uslongI}]. We present a brief discussion of
the collective mode predictions of the Chern Simons approach for 
double-layer systems in
Section~\ref{sec:SMA} of this Chapter.

\section{Fractional Charges in Double-Layer Systems}
\label{sec:frac-charge}

The fractional quantum Hall effect occurs because,
at particular filling factors, electrons
in a partially filled Landau level are able to organize
themselves into such strongly correlated states that the
energy cost of making decoupled particles and holes remains
finite, even in the thermodynamic limit, {\it i.e.}, there is
a charge gap.  It is unusual to have a charge gap due
entirely to electron-electron interactions 
(i.e. in a tranlationally invariant continuum system)
although the example of superconductivity is
 familiar. In the fractional quantum Hall effect, not only
do interactions produce a charge gap, but the free-charges
responsible for the thermally activated dissipation measured
experimentally contain only a fraction of the charge of
an electron.  The fact that sharply-defined
fractional charges occur in the
fractional quantum Hall effect can be understood as a necessary
consequence of the quantization of the Hall
conductance.\cite{leshouches,RBLchap7}  (See the related 
discussion in Section~\ref{sec:field-theory}.) 
It can also be understood in terms of the variational wavefunctions
introduced by Laughlin in his pioneering work on the theory of the
fractional quantum Hall effect.\cite{RBLchap7}  In this section we
discuss the fractionally charged excitations of the $\Psi_{m,m',n}$ states
by generalizing the plasma arguments made by Laughlin for
single-layer systems.

The $\Psi_{m,m',n}$ wavefunctions have the property that
pairs of electrons are excluded from certain relative angular momentum
states.  A low-energy charged state must have a localized excess or
deficiency of charge without destroying the energetically
favorable correlations associated with the relative angular momentum state
exclusions.  For a single-component system, Laughlin suggested that
an accurate approximation to the many-body wavefunction for a state
with a charged hole at the origin could be obtained simply by multiplying
his wavefunctions for incompressible states by the factor $\prod_{i} Z_i$.
For two-component systems this argument has an obvious generalization.
We can produce two, in general different, charged excitations centered on
the origin, by multiplying the orbital many-particle wavefunction by the product
of $Z_i$ for all electrons in one pseudospin state.  This operation is the
variational wavefunction equivalent of introducing an unattached
flux-tube in Chern-Simons theories and the conclusions we reach
below can equally well be obtained by using an algebraically
equivalent argument in that language.  The plasma analogy results
from writing the quantum distribution function, the square of the
many-body wavefunction, as a classical statistical mechanics
distribution function for interacting particles in an external
potential
\begin{equation}
|\psi |^{2} = e^{-U}.
\label{eq:plasanol}
\end{equation}
The classical systems that result are generalized two-dimensional
Coulomb plasmas\cite{forrester} and it is convenient in discussing
them to adopt the convention of using Roman indices for one
component of the plasma and Greek indices for the other component.
With this notation, the trial wavefunctions we consider for charged
excitations are
\begin{equation}
\Psi^A_{m,m',n} \equiv [\prod_{i=1}^{N_R} Z_i ] \Psi_{m,m',n}
\label{eq:holewfa}
\end{equation}
and
\begin{equation}
\Psi^B_{m,m',n} \equiv [\prod_{\alpha=1}^{N_G} Z_\alpha] \Psi_{m,m',n}.
\label{eq:holewfb}
\end{equation}
These trial wavefunctions clearly reduce the density near the origin
without ruining the good correlations in $\Psi_{m,m',n}$.  At first
sight it might appear that only the Roman particle density is reduced
in $\Psi^A_{m,m',n}$ and only the Greek particle density is reduced
in $\Psi^B_{m,m',n}$, but this is not the case in general because of the
correlation factors.

The classical potential energy corresponding to $\Psi^A_{m,m',n}$ is
\begin{eqnarray*}
U^A_{m,m',n} &=& m \sum_{i<j} (-2\ln |Z_{i} - Z_{j}|) +
m' \sum_{\alpha <\beta} (-2\ln
|Z_{\alpha} - Z_{\beta}|) + n \sum_{i,\alpha} (-2\ln |Z_{i} -
Z_{\alpha}|)\\
 && + \sum_{i} \frac{|Z_{i}|^{2}}{2} + \sum_{\alpha}
\frac{|Z_{\alpha}|^{2}}{2} +\sum_{i} - 2 \ln |Z_i|.
\end{eqnarray*}
This potential is a {\it generalized} two-dimensional Coulomb
plasma\cite{forrester} in the sense that the coupling constants outside the
sums in the interaction terms are not constrained to be
the products of charges for the two species, {\it i.e.,} we allow
$ n \ne (m m')^{1/2}$.  This difference results in long-range
interactions in the plasma which depend on the density of each
species separately,
rather than just on the total `charge'.  $U^A_{m,m',n}$
is the potential energy function for a system consisting of
Roman and Greek particles.  All particles have repulsive mutual two-dimensional
Coulomb interactions with coupling constant $m$ between two
Roman particles, coupling constant $m'$ between two Greek particles,
and coupling constant $n$ between a Roman particle and a Greek particle.
All particles are attracted to a
neutralizing background which can be considered to have resulted from
interaction with unit coupling constant with non-responding particles
of uniform charge density $(2 \pi \ell^2)^{-1}$. For $U^A$,
only Roman particles interact with unit coupling constant with
an impurity particle located at the origin.

The charge densities induced
in each species of particles by the impurity can be calculated using
the perfect screening properties which result from the long-range
interactions of the plasma.  Far enough from the impurity the direct
long-range interaction must vanish for each species of particle;
{\it i.e.,} the sum of the impurity charge times its coupling
constant plus the induced charges in each plasma component times
the coupling strength for that plasma component must vanish so that
\begin{equation}
\left(\begin{array}{cc}
  m & n\\
  n & m'\end{array}\right) \left(\begin{array}{c}
    e_{R}^{A}\\
    e_{G}^{A}\end{array}\right) = \left(\begin{array}{c}
      1\\
      0\end{array}\right).
\label{eq:perscreen}
\end{equation}
In Eq.(\ref{eq:perscreen}) $e_{R}^{A}$ is the contribution,
in units of the magnitude of the electron charge, to the
quasiparticle charge from Roman particles and $e_{G}^{A}$ is
the contribution from Greek particles.  Eq.(\ref{eq:perscreen})
can be solved for $e_{R}^{A}$ and $e_{G}^{A}$ and the
total quasiparticle charge $e_{T}^{A} \equiv e_{R}^{A} + e_{G}^{A}$:
\begin{equation}
e_{R}^{A} = \frac{m'}{mm'-n^{2}},\qquad e_{G}^{A} =
\frac{-n}{mm'-n^{2}},\qquad e_{T}^{A} = \frac{m'-n}{mm'-n^{2}}.
\label{eq:fcb1}
\end{equation}
The fractional charges for $\Psi^{B}_{m,m',n}$ differ only
through the interchange of $m$ and $m'$:
\begin{equation}
e_{R}^{B} = \frac{-n}{mm'-n^{2}},\qquad e_{G}^{B} =
\frac{m}{mm'-n^{2}},\qquad e_{T}^{B} = \frac{m-n}{mm'-n^{2}}.
\label{eq:fcb2}
\end{equation}
Fractional charges calculated from these expressions are listed in Table
\ref{table:fc}.

We note that the total charge of what is presumably the
lowest energy charged excitation at each filling factor, has a value $e/q$
where
$q$ is the denominator of the fractional total filling factor.
When the two-components of the incompressible Hall fluid are
correlated, a reduction of charge density in one-component
leads to an increase of charge density in the other component.
The total charge thus tends to consist of partially canceling contributions
from the two-layers.  This cancellation reaches its extreme limit  for the
case where $m =m' = n $, for which the total charge of the
excitation is well defined but its separation into
contributions from separate components cannot be fixed by the
perfect screening requirement on the plasma.  We will see
later that what is behind this behavior is the existence of
long-range-order in the $\Psi_{m,m,m}$ wavefunction.  This long range
order has delivered a bonanza of new physics in two-component
systems, which will be the focus of much of this Chapter.


\section{Collective Modes in Double-Layer Quantum Hall Systems}
\label{sec:SMA}

In this section we will discuss both
intra-Landau-level and inter-Landau-level (cyclotron)
collective (neutral) excitations of the
incompressible ground states whose origin we have explained
in previous sections.  Our discussion is based on the
projected single-mode-approximation, which has proved extremely
useful\cite{GMP,mog,caveat1}
in understanding the nature of the collective mode structure in
single-layer systems.  The projected single-mode approximation
can be appropriate in the strong magnetic field limit
where there is little Landau level mixing in either the
ground state or the low-lying excited states.  Many-body
eigenstates of the system can then be distinguished by
the quantized integer number of units of
$\hbar \omega_c$ by which the
kinetic energy exceeds the minimum value $N \hbar \omega_c/2 $
(where $N$ is the number of particles).
The single-mode-approximation for the collective energy spectrum
follows from the assumption that there is a unique many-body state,
 $|\Psi^n_k\rangle$, with energy $E^n_k$ within
each quantized kinetic energy manifold, which is coupled to the
ground state by the one-body density operator:
\begin{equation}
\rho_k \equiv \sum_i \exp (i {\bf k} \cdot  {\bf r}_i).
\label{eq:cm1}
\end{equation}
[We will employ complex number notation ($k\equiv k_x+ik_y$)
for two-dimensional vectors when
convenient.]

To use the single-mode approximation it is necessary
to separately evaluate contributions to moments of the dynamic
structure factor from transitions involving different numbers of quantized
kinetic energy units.  We write for the dynamic structure factor
\begin{equation}
s(k,\epsilon) = \sum_n s_n(k,\epsilon),
\label{eq:cm2}
\end{equation}
where
\begin{equation}
s_n(k,\epsilon) = \frac{1}{A} \sum_{i} | \langle \Psi_{i,n} |
\rho_k | \Psi_0 \rangle |^2  \delta(\epsilon - E_{i,n}+E_{0}),
\label{eq:cm3}
\end{equation}
and $A$ is the area of the system.
Here $|\Psi_{i,n}\rangle$ is an exact eigenstate of the Hamiltonian
involving $n$ excess quantized kinetic energy units.  In the single-mode
approximation it is assumed that only a single eigenstate contributes to the
sum in Eq.(\ref{eq:cm3}),
\begin{equation}
s_n(k,\epsilon) = \frac{1}{A} |\langle \Psi_{k,n}|\rho_k | \Psi_0 \rangle |^2
\delta(\epsilon - E^n_k + E_0).
\label{eq:cm4}
\end{equation}
Both the matrix element and the energy which appear in this expression 
have physical significance.  The matrix element determines how strongly
one-body external probes
(for example, far infrared or microwave radiation) couple to
the collective excitation.  The energies of the collective modes
can be measured in transmission experiments or in inelastic
light-scattering experiments.\cite{resonantraman}  The thermodynamics
and linear response functions of the system depend qualitatively on
the presence or absence of collective modes whose energies vanish
in the limit of long wavelengths.

Two moments of $S_n(k,\epsilon)$ are relatively easy to evaluate and
we will use these two moments to determine both the matrix element and
the collective mode energy:
\begin{equation}
s_n(k) \equiv \int_{0}^{\infty} d \epsilon\, s_n(k,\epsilon),
\label{eq:cm5}
\end{equation}
and
\begin{equation}
f_n(k) \equiv \int_{0}^{\infty} d \epsilon\, \epsilon \,s_n(k,\epsilon).
\label{eq:cm6a}
\end{equation}
Given $s_n(k)$ and $f_n(k)$, we have  for the matrix element
\begin{equation}
|\langle \Psi^n_k | \rho_k | \Psi_0 \rangle |^2 = A s_n(k),
\label{eq:cm6b}
\end{equation}
and  for the excitation energy
\begin{equation}
\Delta^n_k \equiv E^n_k -  E_0 = f_n(k)/s_n(k).
\label{eq:cm7}
\end{equation}

Since $\rho_{k=0}$ is a constant, and the ground and excited states must be
orthogonal,
 it follows that at long-wavelength
$s_n(k)$ must vanish at least as fast as $k^2$.  Long wavelength
probes, like far infrared radiation, produce\cite{resonantraman} observable
coupling to a long-wavelength collective mode only if $s_n(k) \propto
k^2$ at long-wavelengths.  We will refer to modes for which $s_n(k) \propto
k^2$ as dipole active and those for which $s_n(k)$ vanishes with a higher
power of $k$ as dipole inactive. The quantity
  $f_n(k)$ is proportional to the
product of the square of the matrix-element and the excitation energy
and in analogy with atomic physics we refer to this quantity as the
projected oscillator strength.  We will refer to $s_n(k)$ as the
projected static structure factor.  The usual `f-sum rule', valid with or
without a magnetic field,
states that for parabolic bands with effective mass $m^*$
\begin{equation}
\sum_n f_n (k) = \frac{N}{A} \hbar^2 k^2 / 2 m^*.
\label{eq:cm8}
\end{equation}
To evaluate projected `f-sum rules' we exploit the property that
the single-particle
Hilbert space of a charged particle in a magnetic field can\cite{leshouches} be
considered as the product space of a factor space in which states are
distinguished
by the number of quantized kinetic energy units in the cyclotron orbit
and a factor space for the cyclotron-orbit-center degree of freedom which
exists within each Landau level and is responsible for the macroscopic
Landau level degeneracy.  To separate the dynamic structure factor into
contributions associated with different quantized kinetic energies we
write the kinetic energy operator in the form
\begin{equation}
\rho_{k} = \sum_{n',n} \rho_{k}^{n',n},
\label{deceq:1}
\end{equation}
where
\begin{equation}
\rho_k^{n',n} = \sum_i |n'\rangle_i\langle n|_i G^{n',n}(k) B_i(k).
\label{deceq:2}
\end{equation}
The $G^{n',n}(k)$ are related to Laguerre polynomials\cite{leshouches}, the
sum over $i$ is over particle labels, and $B_i(k)$ is a factor
 coming from the
projection of $\rho_k$ onto a single Landau level and operates on the
intra-Landau-level degree of freedom of  particle $i$.  The
commutators which appear below are evaluated by  using
Eq.~(\ref{deceq:2}) and the identity,
\begin{equation}
B_i(k_1) B_i(k_2) = \exp (k_1^{*} k_2/2) B_i(k_1+k_2).
\label{deceq:3}
\end{equation}
At strong magnetic fields, only
$\rho_{k}^{n,0}$ contributes to $s_n(k)$ and $f_n(k)$.
We will restrict our attention to the cases of intra-Landau-level
($n=0$) and magnetoplasmon ($n=1$) collective excitations.  (Excitation
modes out of the LLL with $n>1$ are never dipole active.)

Using completeness relations of the many-body eigenstates it follows from
the above definitions that
\begin{equation}
s_0(k) = \frac{1}{A}
 \langle \Psi_0 | \rho_{-k}^{0,0} \rho_k^{0,0}|\Psi_0 \rangle
\label{eq:may2a}
\end{equation}
and that
\begin{equation}
f_0(k) = \frac{1}{2A} \langle 
\Psi_0 | [[\rho_{-k}^{0,0},\hat V],\rho_{k}^{0,0}]
|\Psi_0 \rangle
\label{eq:may2b}
\end{equation}
where $\hat V$ is the electron-electron interaction term in the
Hamiltonian.  From general properties of (translation invariant)
many-body eigenstates within the lowest
Landau levels it can be shown\cite{GMP} that for small $k$,
$s_0(k) \propto k^4$ so that the intra-Landau
level collective mode is dipole inactive.  After a somewhat laborious
calculation, involving repeated application of Eq.(\ref{deceq:3}),
the $n=0$ projected oscillator strength can be expressed
in terms of $s_0(k)$ and one finds the result that $f_0(k) \propto
k^4$, independent of any details of the electron-electron interaction
or the ground state wavefunction,
so that the magnetoroton intra-Landau-level collective modes have
a gap at long-wavelengths:
\begin{equation}
\lim_{k \to 0} \Delta^0_k \ne 0.
\label{eq:gap}
\end{equation}
On the other hand an elementary calculation
based on Eq.(\ref{deceq:3}) and the strong-field-limit
assumption that the ground state lies entirely within the
lowest-Landau-level subspace of the full Hilbert space, implies that
\begin{equation}
s_1(k) = \frac{N}{A} \frac{k^2 \ell^2}{2} \exp (- |k \ell|^2/2 ).
\label{eq:may2c}
\end{equation}
Similarly\cite{mog}
\begin{equation}
f_1(k) = \hbar \omega_c s_1(k) +  \frac{1}{2A} \langle \Psi_0 |
[[\rho_{-k}^{0,1},\hat V],\rho_{k}^{1,0}] |\Psi_0 \rangle,
\label{eq:may2d}
\end{equation}
and the second term on the right-hand-side of Eq.(\ref{eq:may2d}) can be
shown to vanish as $k^3$ for small $k$.  The magnetoplasmon mode
is dipole active and, in the long-wavelength limit, completely
exhausts the full f-sum rule.  For a single-layer, the magnetoplasmon
mode is the only dipole-active mode and its energy is not shifted from
$\hbar \omega_c$ by electron-electron interactions.
These behaviors result from the conservation of
particle number and invariance under translation: a long wavelength
electromagnetic field couples only to the cyclotron motion of the
center of mass of the system.\cite{kohn}

We are now prepared to
discuss how these results are altered in double layer systems.
We restrict our attention to the case where the two 2DEG's are
identical and tunneling between them may be neglected.  In this case
the number of electrons in each layer is a good quantum number and
collective modes corresponding to the sum and difference of the
density oscillations in the two layers decouple.
To generalize the projected
single-mode-approximation to the double-layer case
we evaluate separately projected oscillator strengths for
both sum (in phase) and difference (out of phase) modes.
We will find that the behavior of the sum modes for double-layer
systems is similar to the behavior of the modes of a single-layer system
while the behavior of the difference modes departs from this pattern.
For the difference modes,
both $n=0$ and $n=1$ modes are dipole active.  The
$n=1$ mode is shifted from $\hbar \omega_c$ and the
$n=0$ mode {\it usually} has a finite energy as $k \to 0$.
An exception occurs for those ground state wavefunctions which have a
type of
long range order which we have not yet made explicit.  This long range
order is associated with a broken symmetry ground state.
These broken symmetry ground states are, arguably, responsible for
the most surprising and appealing new physics which is introduced on
going from one-component to two-component fractional quantum Hall
systems and we will have much more to say about them later in this Chapter.
The difference in behavior between sum and difference modes is
due to the fact that the Hamiltonian is not invariant under
relative translations of the two layers.

The Hamiltonian of the double-layer system in the absence of
interlayer tunneling may be written in the following form which is
convenient for calculations:
\begin{eqnarray}
H & = & \hbar\omega_c \sum_i [a_i^{\dagger}(L) a_i(L) +
a_i^{\dagger}(R) a_i(R)] + \nonumber \\
  && {\textstyle{ 1 \over 2}}  \sum_q [V_q^{LL} \rho_q(L)\rho_{-q}(L)+
V_q^{RR} \rho_q(R)\rho_{-q}(R)+ 2  V_q^{LR} \rho_q(L)\rho_{-q}(R)],
\label{eq:1}
\end{eqnarray}
where $a_i(L)$ is the Landau level lowering\cite{leshouches}
operator for particle $i$
in the left (L) layer, $V_q^{LL} = V_q^{RR}$ is the intra-layer
Coulomb interaction, $V_q^{LR}$ is the inter-layer Coulomb
interaction, and $\rho_q(X)$ is the density operator for layer X.
[For explicit calculation we ignore the finite thickness of the
2D layers so that $V_q^{LL} = 2 \pi e^2 /(\epsilon q)$ 
and $V_q^{LR}=\exp (-q d) V_q^{LL}$.]  
In Eq.~(\ref{eq:1}) the  Hamiltonian includes infinite
constant terms corresponding to the self-interaction of each electron
in the system.  It is  convenient to retain these terms so that the
interaction terms  in the Hamiltonian can be expressed in terms of
density operators.   Since, in all subsequent calculations, the
Hamiltonian enters only in commutators, these non-physical constant
terms never contribute.  The operators which generate the sum and
difference collective modes are
$\Omega_k^{n\pm} =  [\rho_k^{n0}(L) \pm \rho_k^{n0}(R) ]/ \sqrt{2}$.
With the Hamiltonian expressed in terms of density operators,
a laborious but direct calculation makes it possible to
express $f_k^{n\pm}$ and $s_k^{n\pm}$ in terms of
$s_k^{n\pm}$ using Eq.(\ref{deceq:3}).  Explicit expressions for
the collective mode energies and the coupling matrix elements are
given elsewhere.\cite{zhang-macd-collective,renn2}
  Here we only comment on some
of the physically interesting conclusions of these calculations.

As mentioned above the energy of the $n=1$ sum mode
approaches the cyclotron energy $\hbar\omega_c$ in the $k \to 0$
limit, in agreement with Kohn's theorem.  (Since interactions are
invariant under simultaneous translations in both layers, the proof for
a single-layer system\cite{kohn} trivially generalizes to the case of
the two-layer in-phase mode.) However the energy of the $n=1$ difference
mode is shifted from the cyclotron energy in the long
wavelength limit where it is given by the following expression:
\begin{equation}
\Delta^{1-}(k=0) = \hbar\omega_c - \int {d^2 q \over (2\pi)^2}
 q^2 V_q^{LR} h^{LR}(q).
\label{eq:9}
\end{equation}
Here $h^{LR}(q)$ is the Fourier transform of the inter-layer pair correlation:
\begin{equation}
h^{LR}(q) = \frac{1}{N}\langle \Psi_0 | \rho_q(L) \rho_{-q}(R) |
 \Psi_0 \rangle,
\label{eq:pcf}
\end{equation}
and
\begin{equation}
h^{LL}(q) =
\frac{1}{N}\langle \Psi_0 | \rho_q(L) \rho_{-q}(L) | \Psi_0 \rangle -1 ,
\label{eq:pcfp}
\end{equation}
where $N$ is the number of particles per layer.
If the layers are uncorrelated,
$h_{LR}(q) = h_{LL}(q) = 0$ and $\Delta^{1-}(k=0) = \hbar \omega_c$.
For correlated layers $h_{LR}(q)$ tends to
be negative, at least at small $q$ since the density in the left
layer will tend to be reduced when the density in the right layer is
increased, and we can expect that $\Delta^{1-}(k=0) > \hbar \omega_c$.
In fact, it is possible to prove that
interactions (of either sign!) always increase the
frequency of this mode as we mention below.

By expanding the expression for $ f_k^{0\pm}$ at small $k$
it can be shown that $ f_k^{0+} \sim k^4$ whereas
\begin{equation}
f_k^{0-} = - \frac{N}{A} (k^2/2)  \sum_q q^2 V_q^{LR} h^{LR}(q) + {\cal O}(k^4).
\label{eq:12}
\end{equation}
The $n=0$ difference mode is dipole active.
It is interesting that for the difference mode the interaction
contributions to the dipole ($\propto |k|^2$) portions of the
$n=0$ and $n=1$ oscillator strengths are
identical. By definition $f_k^{0-}$ is positive definite so that
$\Delta^{-}(k=0) - \hbar \omega_c$ must also be positive.  In the
single-mode approximation, the  $n=1$ (cyclotron) difference mode is always
shifted to higher energy by electron-electron interactions.  The
situation is similar to that for the effect of disorder on the
vibration modes of the Wigner  crystal at strong magnetic fields where
pinning of the crystal shifts both intra-Landau-level and
inter-Landau-level modes upward by the same amount\cite{macdunpub}.

To determine whether or not the $n=0$ mode is gapped,
it is necessary to determine how $s^{0\pm}_k$ behaves at small $k$.  From
general properties of wavefunctions in the lowest Landau level
it is possible\cite{zhang-macd-collective} to conclude that\cite{exceptnu1}
$ s^{0+}_k \sim k^4$ , whereas the behavior of the
$n=0$ difference mode structure factor depends on the difference between
inter-layer and intra-layer correlation functions.
It is possible to prove\cite{zhang-macd-collective}
that $s^{0-}_k \sim k^2$ at small $k$
provided that intra-layer and inter-layer correlation functions
separately vanish at large spatial separations.  In general it is easy
to show from the plasma arguments outlined in
Section~\ref{sec:frac-charge} that
this is a property of the $\Psi_{m,m,n}$ wavefunctions.  However
an exception occurs for $n=m$.  In this case there is no distinction between
the effective plasma interactions of particles in the same layer and
particles in different layers.  The weighting of particle configurations
depends only on the {\em total} charge densities of the two-layers and only
correlations in the {\em total} charge densities of the two-layers go to
zero at large distances.
For the broken symmetry state
$\Psi_{m,m,m}$
it is possible\cite{zhang-macd-collective} to show
explicitly that $s^{0-}_k = \exp(-|k \ell|^2/2)$ which goes to a constant
for $k \to 0$.  The SMA collective mode then goes like $k^2$ for this
special case {\it if} the ground state is approximated by $\Psi_{m,m,m}$.
(The italics above are pregnant as we will see later.)  For the
 $\{m,m,m\}$ states, the
$n=0$ difference mode (intra-Landau level)
is gapless at long-wavelengths.  It is often
the case that gapless collective modes can be identified as Goldstone
modes associated with a broken symmetry in the ground state,  and that is
indeed the case here, as we shall see.

We emphasize that the results discussed above follow from general sum rules and
are independent of the approximate many-body wavefunction ($\Psi_{m,m',n}$)
in terms of which we have framed our discussion so far.
Since $ f^{0+}_k \sim k^4$ and $ f^{0-}_k \sim k^2$
independent of the long wavelength  behavior of
$s^{0\pm}_k$, it follows quite generally that the $n=0$ sum
mode has a gap and that the $n=0$ difference
mode has a gap except in the case where long-range order is present
which results in correlation functions which do not vanish at large
distances.
In Fig.~(\ref{fig:cm1}) we show
results obtained for the collective mode energies of
a double-layer system with a total Landau level
filling factor $\nu_T = 1/2$ and a layer separation $d = 1.5 \ell$,
close to the effective layer separation value for which novel
double-layer fractional Hall effects have recently been
observed\cite{exp}.  Numerical calculations\cite{nu1/2refs,chakra87} have
established that the ground state at this value of
$d/\ell$ is accurately approximated by the
$\{m,m,n\}=\{3,3,1\}$ Halperin\cite{halperinz1}
wavefunction and we have used the correlation functions\cite{rm,baps}
of that wavefunction to evaluate the oscillator strengths and
structure factors. For $k \to 0$ the $n=1$ sum mode
(the Kohn mode) is unshifted by interactions while the $n=1$ difference
mode is shifted to higher energies as discussed above. The shift,
which is a direct measure of inter-layer correlations,
should\cite{caveat2} be observable in cyclotron resonance experiments
in double-layer systems.  Note also that both sum and difference $n=0$
modes have a finite gap as expected from the above discussion.

We are now able to compare our results for the collective modes with
the Chern-Simons Landau-Ginzburg (CSLG)
 theory of the double-layer system\cite{Wen-Zee_double,ezawa,ezawa_95}. In the
CSLG random-phase-approximation, the sum and difference density
response functions are given by:
\begin{equation}
\rho_{+} (\omega, q) = \frac {N q^2 /A m^*} {\omega^2 -
\omega_{+}^2}  \ \ \ , \ \ \ \rho_{-} (\omega, q) = \frac {N
q^2 /m^*A } {\omega^2 - \omega_{-}^2},
\label{eq:14}
\end{equation}
where the collective mode frequencies are given by 
$\omega_{+} = \omega_c = eB/m^* c$, $\omega_{-} = \omega_c
(m-n)/(m+n)$ and $m^*$ is the effective band mass of the electrons.
For the single-layer case the CSLG random-phase-approximation predictions
are correct for the dipole active mode and we might have expected the
same to be true in double-layer systems.
The sum mode for double-layer systems can be clearly identified with the Kohn
mode\cite{lz}.  However there are difficulties in
identifying the density difference modes in this theory.  From
Eq.~(\ref{eq:12}) we see that the $n=0$ difference mode is dipole active
and should have a dipole oscillator strength proportional to $V^{LR}
h^{LR}$.  One
might therefore be tempted to identify $\omega_{-}$ with the
$n=0$ difference mode. Then, for the case of the $\{m,m,m\}$ state
random-phase-approximation, $\omega_{-} = 0$, and it would then be
tempting to identify this mode with the gapless $n=0$
Goldstone mode.  Unfortunately, the single density-difference mode
calculated within the double-layer CSLG theory
saturates the full dipole oscillator strength $N q^2/m^* A$. This
is not acceptable since an excitation within the lowest Landau level
can not contain explicit dependence on the band mass $m^*$.  The
second possibility is to interpret the difference mode obtained in
Eq.~(\ref{eq:14})
as the $n=1$ difference mode.  In this case one is
faced with the difficulty that the mode energy is shifted downwards
from the cyclotron energy by an amount proportional to $\omega_c$,
whereas the SMA calculations show that it should be shifted upwards by
an amount proportional to the interlayer Coulomb energy.  It has
been suggested\cite{inprogress} that these difficulties can be resolved by
including the mixing of the vortex excitations with the gaussian
fluctuations in the CSLG theory.  Similar difficulties arise in
the fermion-Chern-Simons theory of the single-component $\nu =1/2$
state and can be resolved in that case by taking sufficient care with the
Landau parameters of the composite-fermion Fermi liquid.\cite{halperinchap}


\section{Broken Symmetries}
\label{sec:broken-symm}

Ferromagnetic states break spin rotation symmetry since they
are defined by an order parameter $\left\langle {\bf S}\right\rangle$
which gives the
magnitude and orientation of the magnetization.  For the SU(2) symmetric
case with no Zeeman term, this orientation is arbitrary.  As we will see
below, the case of a double layer system is described by a pseudospin with
easy-plane anisotropy [U(1) or XY symmetry].  Here the magnetization vector
is forced to lie in the XY plane in the ground state.  The origin of
ferromagnetism in all these systems is the Coulomb interaction just as it is
for itinerant ferromagnets like iron.  Exchange effects are particularly
crucial in a 2DEG in a large magnetic field because the kinetic energy is
quenched into highly degenerate Landau levels.  It is advantageous to
follow Hund's rule and maximize
the spin in order to make the spatial wave function fully antisymmetric,
thereby lowering the Coulomb energy.  Since the Landau level is degenerate,
this spin alignment can in some cases be complete since
it costs no kinetic energy as it does in iron.\cite{foot:hand}

Before considering the physical consequences of this broken symmetry, let
us return to Table~\ref{table1} 
to consider how the total spin quantum number $S$ for
a state can be determined.  A portion of our discussion here follows that of 
Ref.[\citen{macdsurface}].
As already mentioned, states of the form $\{m,m,m\}$ are fully ferromagnetically
aligned and have total spin $S=N/2$.  To derive the spin quantum numbers
for the other states in Table~\ref{table1} we write
\begin{equation}
{\bf S}_T^2 \Psi [Z:\chi ]=A [\Phi' [Z] \alpha_1\cdots \alpha_{N_\uparrow}
\beta_{[1]}\cdots \beta_{[N_\downarrow]}],
\end{equation}
and use the fact that
\begin{eqnarray}
{\bf S}_T^2 &=& (S^z)^2+\frac{1}{2}(S^+S^-+S^-S^+)\nonumber\\
 &=& (S^z)^2+\frac{N}{2}+\frac{1}{2} \mathop{{\sum}'}_{k,l}
(S^+_{k}S^-_{l}+S^-_{k}S^+_{l}). \label{eq50}
\end{eqnarray}
We see that
\begin{eqnarray}
 && \Phi' [Z]=\left[\left(\frac{N_\uparrow
-N_\downarrow}{2}\right)^2+\left(\frac{N}{2}\right)\right] \Phi
[Z]\nonumber\\
 && \phantom{\Phi'}-\sum_{i=1}^{N_\uparrow} \sum_{j=1}^{N_\downarrow}
e(i,[j]) \Phi [Z]=S (S+1) \Phi [Z], \label{eq60}
\end{eqnarray}
where $e(i,k) \Phi (\cdots ,Z_i,\cdots , Z_k,\cdots )=\Phi (\cdots
,Z_k,\cdots ,Z_i,\cdots )$ is a label exchange operator.
In Eq.~(\ref{eq50}), $S^\pm\equiv
\sum_kS^{\pm}_{k}$ and $S^{+}_{k}$ and $S^{-}_{k}$
are spin raising and lowering
operators,
 and the prime on the sum in
Eq.~(\ref{eq50}) indicates that $k=l$ is excluded.  In Eq.~(\ref{eq60})
$S$ is the total-spin quantum number.

To identify Halperin $\{m,m',n\}$ states which are total spin eigenstates,
we start with states where $m=1$, $n=0$ and $m'$ has any odd value.
Since $\prod_{i<j} (Z_i-Z_j)$ is a Vandermonde determinant these orbital
wave functions have the up-spin Landau level full (see table~\ref{table1}).
They are eigenstates of $S^z$ with eigenvalue
\begin{equation}
(N_\uparrow
-N_\downarrow )/2=N_\downarrow (m'-1)/2=N(m'-1)/2(m'+1)\equiv S_{m'}.
\end{equation}
Moreover $S^+\Psi_{1,m',0}\equiv 0$ since the up-spin Landau level is
already full and there are no wave functions with a larger value of
$N_\uparrow$.  It follows that $\Phi_{1,m,0} [Z]$ satisfies
Eq.~(\ref{eq60}) with $S=(N_\uparrow -N_\downarrow)/2$ and hence that
\begin{equation}
\sum_{i=1}^{N_\uparrow} \sum_{j=1}^{N_\downarrow} e(i,[j],) \Phi_{1,m',0}
[Z]=N_\downarrow \Phi_{1,m',0} [Z]. \label{eq100}
\end{equation}
(This result can also be established by an explicit algebraic proof).
Since $e(i,[j]) Q[Z] \Phi [Z]=Q[Z] e(i,[j]) \Phi [Z]$ for any symmetric
polynomial $Q[Z]$, we have from Eqs.~(\ref{eq100}) and (\ref{eq60}) that
\begin{equation}
 {\bf S}_T^2 \Phi_{1+2p,m'+2p,2p}[Z:\chi]
 =S_{m'}(S_{m'}+1) \Phi_{1+2p,m'+2p,2p}\left[Z:\chi\right].
\end{equation}
In addition, it follows from Eq.~(\ref{eq60}) that ${\bf S}_T^2 \Psi
\left[Z:\chi\right]=N(N/2+1)/2 \Phi\left[Z:\chi\right]$
for any completely antisymmetric
function $\Phi [Z]$, and in particular for generalized Laughlin states
with $m=m'=n$.  These states are merely the $S^z=0$ members of the set
of the $(N+1)$ fully polarized Laughlin states which are degenerate in
the absence of the Zeeman term.

We return now to the question of the physical consequences of the
spontaneously broken symmetry of ferromagnetic states.  We will focus
initially on the SU(2) invariant case of `real' spins with zero $g$ factor.
Consider a state with $\nu=1$ and all spins up.  Because the up Landau level
is maximally  filled, the Pauli principle forces us to flip a spin if
we are to move an electron to create a pair of charged excitations.
This is illustrated in Fig.[\ref{fig:spin-flip} a] and shows that
spin and charge are intimately connected in this case.

Transport
dissipation measures the thermal activation of charged excitations.
In the absence of interactions, the energy cost of charged excitations is
zero and there will be no $\nu=1$ quantum Hall plateau because there is no
gap.  In the presence of Coulomb interactions a flipped spin particle-hole
pair causes a loss of exchange energy of\cite{uslongI}
\begin{equation}
E_0 = \sqrt{\frac{\pi}{2}} \frac{e^2}{\epsilon\ell},
\label{eq:E_0}
\end{equation}
where $\epsilon$ is the bulk dielectric constant.
This occurs because the electron spatial wave function is no longer
perfectly antisymmetric.
This energy gap is quite large ($\sim 150K$ at $B=10$T)
and is vastly larger than the bare Zeeman splitting.  Hence Coulomb
interactions and the associated ferromagnetism play a dramatic role
in producing the charge gap at $\nu=1$ (and
because of the spontaneous magnetic ordering,
will continue to do so, even if $g$ is strictly zero).

The exchange energy cost of particle-hole pair excitations  is so large
that it is worth searching for some modified form of the excitation
which is less costly.
A prescient analysis of smooth spin textures by
Sondhi, et al.\cite{sondhi} yielded the exciting
idea that `skyrmion'\cite{kaneandlee} like
spin textures (shown in cross section in Fig.[\ref{fig:spin-flip} b])
can have relatively low energy and
carry fermion number proportional to their topological charge
(Pontryagin index)\cite{sondhi,uslongI}
\begin{equation}
\Delta N=-\frac{\nu}{8\pi}\>
\int{d^2{\bf r}}\, \epsilon_{\mu\nu}\> {\bf m}({\bf r})\cdot
\left[\partial_\mu {\bf m}({\bf r})\times \partial_\nu {\bf m}({\bf
r})\right],
\label{eq3.245}
\end{equation}
where ${\bf m({\bf r})}$ is the unit vector field representing the local
spin orientation. The fermion number $\Delta N$
is an integer multiple of $\nu$ because it is the number of times
the unit sphere is wrapped around by the order parameter.  That is,  it is
the winding number of the spin texture.\cite{fradkin_book}
For the Laughlin parent states
$\nu =1/m$,  elementary spin-textures
carry the same fractional charge
as the quasiparticles discovered by Laughlin\cite{RBLchap7}
for spinless electrons.  As we discuss below, the fact that the
charges are the same follows from very general considerations.
Actually, the spin texture states we have defined must contain
precisely the same number of particles as $|\psi_0 \rangle$ since
the spin-rotation operator does not change the total electron
number.  However the spin-density may contain a number of well-separated
textures with well-defined non-zero topological charge densities and
hence well localized charges; only the net charge in the spin-texture
states defined above will be zero.  The system clearly has states
with locally non-zero net charge in the spin textures.

A simple variational wave function for a skyrmion of size $\lambda$
centered on the origin
and carrying $p$ units of topological charge is given by
\begin{equation}
\psi_\lambda = \prod_j \left(\begin{array}{c} Z_j^p \\ \lambda^p \\ \end{array}
\right)_j \Phi_{mmm},
\label{eq:skyrmicro}
\end{equation}
where $\Phi_{mmm}$ is defined in Eq.(\ref{eq:halperin-multi-psi}),
$()_j$ refers to the spinor for the $j$th particle, and the variational
parameter $\lambda$ is
a fixed length scale.  This is a skyrmion because
the $\hat x-\hat y$ component of the spin has a vortex centered on the 
origin and the $\hat z$-component 
is purely down at the origin
(where $Z_j = 0$) and purely up at infinity (where $|Z_j| \gg
\lambda$) as shown in Fig.[\ref{fig:spin-flip}].
 The parameter $\lambda$ is simply the size scale of the
skyrmion\cite{sondhi,rajaraman}.   Notice that in the limit $\lambda
\longrightarrow 0$ [where the continuum effective field theory
is invalid (see Section~\ref{sec:field-theory}),
but this microscopic wave function is still sensible]
we recover a fully
spin polarized filled Landau level with  $p$ Laughlin quasiholes
at the origin.  Hence the number of flipped spins associated with
the presence of the skyrmion
interpolates continuously from zero to infinity as $\lambda$ increases.

In order to analyze the skyrmion wave function in Eq.(\ref{eq:skyrmicro}),
 we use the Laughlin plasma analogy.\cite{RBLchap7}
In this analogy the norm of $\psi_\lambda$,
 $Tr_{\{\sigma\}}\int D[z] |\Psi[z]|^2$
is viewed as the partition function of a Coulomb gas.
In order to  compute the density distribution
we simply need to take a trace over the spin (we specialize here to the
case $p=1$ for simplicity)
\begin{equation}
Z=\int D[z]\, e^{\frac{2}{m}\left\{m^2\sum_{i>j} \log|Z_i
-Z_j| + \frac{m}{2}\sum_k \log(|Z_k|^2+
\lambda^2) - \frac {m}{4}  \sum_k |Z_k|^2\right\}}.
\label{eqm20}
\end{equation}
This partition function describes the usual logarithmically interacting
charge $m$ Coulomb
gas with uniform background charge plus
a spatially varying impurity background charge $\Delta\rho_b(r)$,
\begin{equation}
\Delta\rho_b(r)\equiv-\frac {1}{2\pi} \nabla^2 V(r)=
-\frac{\lambda^2}{\pi(r^2+\lambda^2)^2},
\label{eqm30}
\end{equation}
\begin{equation}
V(r)=\frac{1}{2}\log(r^2+\lambda^2).
\label{eqm40}
\end{equation}

For large enough scale size $\lambda \gg \ell$, local neutrality of the
plasma\cite{jasonho}
implies that the excess electron number density is precisely
$\frac{1}{m}\Delta\rho_b(r)$, so that Eq.(\ref{eqm40})
is in agreement with the standard result for the
topological density,\cite{rajaraman} and the skyrmion carries
electron number $1/m$ (for $p=1$) and $p/m$ in general.  These objects
are roughly analogous to the Laughlin quasi-hole.  Explicit wave functions
for the the corresponding quasi-electron objects (`anti-skyrmions')
are more difficult to write down just as they are for the Laughlin
quasi-electron due to the analyticity constraint.\cite{RBLchap7}


Sondhi, et al.\cite{sondhi} have shown that for the case of pure
Coulomb interactions (i.e.,
with no finite inversion layer thickness corrections) the  optimal
skyrmion configuration costs precisely half the energy of the simple
spin flip.\cite{precise_def}
  The reason for this is simply that the skyrmion keeps the
orientation of spins close to that of their neighbors and so loses less
exchange energy.   This will be discussed in more detail from a field
theoretic point of view in Section~\ref{sec:field-theory}.

 Optical\cite{skyrmion_expts} and standard transport\cite{skyrmion_expts2}
 experiments
 show that the charge excitation gap is
indeed much larger than would be expected if interactions were 
neglected and has approximately the
correct Coulomb scale, although there is not yet
precise agreement between the observed gap and the best estimates
including the finite $g$ factor.\cite{sondhi,fertig}  The main source
of error is probably neglect of finite thickness effects.
Calculations including finite thickness corrections do not exist at present.
Quantum fluctuation effects (i.e., corrections to Hartree-Fock)
may also be important.
It should be noted that the uniform ground state of the $\nu=1$ ferromagnet
does not have quantum fluctuations (Hartree-Fock is exact here).  However
for finite $g$ factor, the length scales associated with the skyrmion are
small and quantum corrections could well become important.

The idea that skyrmions
are the lowest energy excitations has received very strong
and unequivocal support from
numerical simulations which show that quite remarkably, adding a single
electron to a $\nu=1$ system (with $g=0$)
suddenly changes it from a fully aligned
ferromagnet ($S=N/2$) to a spin singlet ($S=0$) due to the formation of
a skyrmion.\cite{rezayi,sondhi}  (It should be noted that this occurs in
the spherical geometry.  Things are slightly more complicated on the
torus.\cite{uslongI})

The notion of charges being carried by skyrmion textures has received
additional dramatic experimental confirmation in recent optically pumped
NMR measurements by Barrett et al.\cite{barrett_prl,barrett_science}
Their
Knight shift measurements (see Fig.[\ref{fig:knight-shift}])
indicate that the electron gas
spin polarization has a maximum at filling factor $\nu=1$ and falls off
sharply on each side.  The rate of fall off indicates that each charge
added or removed from the Landau level turns over about 7 spins.  This is
consistent with the charge being carried by skyrmions of finite size.
The size of the skyrmion is determined by a competition between the Zeeman
coupling which wants to minimize the number of flipped spins and the
Coulomb self-energy which wants to expand the skyrmion to spread out the
excess charge density over the largest possible area.  As mentioned above,
Rezayi found numerically that a single charge entirely destroys
the spin polarization of a system (if $g=0$).  Using an effective
 field-theory approach,
Sondhi et al.\cite{sondhi}
have estimated the skyrmion size in the regime of small $g$ factor.
Microscopic Hartree-Fock calculations expected to be more accurate in the
physically accessible Zeeman energy regime have been performed by
Fertig et al.\cite{fertig}  These estimates are roughly
consistent with the experimental value of 7 spins per unit of charge.
It should be noted that for finite $g$ factor, the energetic advantage of
the skyrmion over the simple flipped spin is considerably reduced.\cite{fertig}
Wu and Sondhi have shown that in higher Landau levels, skyrmions have
higher energy than other charged excitations.\cite{wu-shivaji}

For the SU(2) symmetric case discussed in this
section, the existence of a vector order parameter
$\left\langle {\bf S}\right\rangle$ is
in some sense trivial because the magnetization commutes with the
Hamiltonian.  For the case of double layer systems we will see that the
the pseudospin Hamiltonian has only U(1) symmetry and the fact that
a non-zero expectation value of $\left\langle {\bf S}\right\rangle$
appears is highly non-trivial.  We will discuss this in more detail in
Section~\ref{sec:field-theory}.


\section{Field Theoretic Approach}
\label{sec:field-theory}

In this section we study the ferromagnetic
broken symmetry ground state and its
excitations from  the point of view of (quantum)
 Ginzburg-Landau effective field theory.
We will begin with the SU(2) invariant case of `real' spin and then move
on to the pseudospin analogy in double-layer systems.  We will give here an
introductory qualitative discussion of the physics. Part of our
discussion follows fairly closely the presentation in Ref.[\citen{uslongI}].
 The technical details of the calculations can be found there and
elsewhere.\cite{sondhi,uslongII,fertig}

The standard first step in this procedure is always to identify the slowly
fluctuating order parameter field,
which in this case we have reason to believe is the local magnetization.
We believe that on long length scales the (coarse-grained) magnetization
fluctuates very slowly, because we know that the zero-wavevector component
of the spin operator (i.e., the total spin) commutes with the Hamiltonian
and so is a constant of the motion.  We will focus on slow tilts of the
spin orientation ignoring variations in the magnitude of the
coarse-grained magnetization, and so define the order in terms of a
local unit vector field ${\bf m}({\bf r})$.

General symmetry arguments can now be used to deduce the form of the
Lagrangian.  We can not have any terms that break spin rotational symmetry
and thus the leading term which is an allowed scalar is
\begin{equation}
E = \frac{1}{2} \rho_{\rm s}
\int d^2r [\nabla {m}^\mu({\bf r})]\cdot[\nabla {m}^\mu({\bf r})],
\end{equation}
where $\rho_{\rm s}$ is a phenomenological spin stiffness coefficient
and the energy is relative to the ground state energy.
It expresses the cost due to loss of Coulomb exchange energy when the
spin orientation varies with position.  For the
SU(2) invariant $\nu=1$ case, this
stiffness may be computed exactly.\cite{uslongI}
For
Coulomb interactions (with no finite thickness corrections)
this calculation yields
\begin{equation}
\rho_{\rm s} = \frac{e^2/(\epsilon \ell)}
{16\sqrt{2\pi }}.
\end{equation}
In two dimensions the stiffness has units of energy and
is approximately 4 K at a field of 10 T.
Numerical estimates for $\nu=1/3$ yield a value which is about 25 times
smaller.\cite{uslongI}

As usual, there is a linear time derivative term
in the Lagrangian which can be deduced from the
fact that each spin precesses under the influence of its local exchange
field.  Equivalently we may note that when the orientation of a spin
is moved around a closed loop, the quantum system picks up a 
Berry's phase\cite{berry}
proportional to the solid angle $\Omega$ enclosed by the path $\omega$
of the tip of the spin
on the unit sphere as shown in Fig.[\ref{fig:berry}].
Noting that a charged particle
moving on the surface of a unit sphere with a magnetic monopole at the
origin also picks up a Berry's phase proportional to the solid angle subtended
by the path\cite{berry} we may express the Berry's phase for a spin $S$ as
\begin{equation}
e^{i\gamma} = e^{i\Omega S} = e^{i\oint_\omega d{\bf m}\cdot
{\bf A}({\bf m})},
\end{equation}
or equivalently
\begin{equation}
e^{i\gamma} =  e^{i\int dt \frac{\partial {\bf m}}{\partial t}
\cdot {\bf A}({\bf m})},
\end{equation}
where ${\bf A}({\bf m})$ is the vector potential of a unit
monopole\cite{sondhi,fradkin_book,uslongI}
at the center of the sphere evaluated at the point ${\bf m}$.  That is,
$\nabla_{\bf m}\times {\bf A} = {\bf m}$.  This phase is correctly
reproduced in the quantum action by adding the following total derivative
term to the Lagrangian for the spin
\begin{equation}
L_1 = S \frac{\partial {\bf m}}{\partial t} \cdot {\bf A}({\bf m}).
\end{equation}
Using the fact that the electronic density is $\nu/2\pi\ell^2$,
the analog for the present problem of a large collection of
electrons with $S=1/2$ at filling factor $\nu$ may be  simply
written
\begin{equation}
{\cal L}_1 =  \frac{\nu}{4\pi\ell^2}\int d^2r
 \frac{\partial {\bf m}}{\partial t} \cdot {\bf A}[{\bf m}],
\end{equation}
which yields the Lagrangian
\begin{equation}
{\cal L} =  \frac{\nu}{4\pi\ell^2}\int d^2r
 \frac{\partial {\bf m}}{\partial t} \cdot {\bf A}[{\bf m}]
-  \frac{1}{2} \rho_{\rm s}
\int d^2r [\nabla {m}^\mu({\bf r})]\cdot[\nabla {m}^\mu({\bf r})].
\end{equation}

We shall see shortly that higher gradient terms can be unexpectedly
significant, but this Lagrangian is adequate to recover the correct
spin wave collective mode.
Taking the spins to be aligned in the ${\bf\hat z}$ direction and
looking at small transverse oscillations at wave vector $\bf q$ we obtain
from this Lagrangian the following equation of motion
\begin{equation}
\frac{d{\bf m}_{\bf q}}{d t} = \frac{4\pi\rho_{\rm s}q^2}{\hbar\nu}{\bf
{\hat z}\times m_q}.
\end{equation}
This yields the dispersion relation
\begin{equation}
\hbar\omega = \frac{4\pi\rho_{\rm s}}{\nu} q^2,
\end{equation}
which agrees with the long-wavelength limit of 
exact results obtained by a variety of means.\cite{kallin,rasolt}

At this point we have expanded the Lagrangian to lowest order in
gradients and we have correctly found the neutral collective spin
wave modes.  Their dispersion is quadratic in wave vector just as it
is for the Heisenberg ferromagnet on a lattice.  However here we
have an itinerant magnet and
we have so far seen no sign of the charge degrees of freedom.  It turns
that we have to go to higher-order gradient terms in the action to
see charged objects.

We have already seen in the discussion of Fig.[\ref{fig:spin-flip}]
in Section~\ref{sec:broken-symm},
that for a filled Landau level,
the Pauli principle forces there to be a connection between charge
excitations and flipped spins.  It turns out
that the existence of a finite Hall conductivity
in this itinerant magnet causes smooth spin textures to carry charge
proportional to their topological density.
One can derive this result from a Chern-Simons
effective field theory,\cite{sondhi,uslongI}
or from microscopic
considerations involving the fact that the spin density and charge density
operators do not commute when projected onto the lowest Landau
level,\cite{uslongI} or from {\em macroscopic} considerations connecting
the Berry phase term to the Hall conductivity.\cite{kagome,uslongI}
The latter is the least
technical and the most instructive so we shall pursue it here.

Imagine that the order parameter of the ferromagnetic system is distorted
into a smooth texture as illustrated in Fig.[\ref{fig:texture}].  As an
electron travels around in real space along  a path $\partial\Gamma$
which is the boundary of the region $\Gamma$, the spin is assumed
to follow the orientation of the local exchange field ${\bf b}({\bf r})$
and hence traces
out a path in spin space labeled $\omega$ in the (schematic)
illustration in Fig.[\ref{fig:berry}].    That is, given any
sufficiently smooth spin
texture, we can write a Hartree-Fock like Hamiltonian for the
electrons which will reproduce this texture self-consistently
\begin{equation}
H = \sum_{j=1}^N {\bf b}({\bf r})\cdot {\bf S}_j.
\end{equation}
If we drag one electron around in real space along
$\partial\Gamma$ and its spin
follows the local
${\bf b}({\bf r})$ adiabatically (which we expect since the exchange energy
is so large) then the electron will acquire a Berry's phase
$\Omega/2$ where $\Omega$ is the solid angle subtended by the region
$\omega$ shown in Fig.[\ref{fig:berry}].

In addition to the Berry's phase from the spin, the electron will acquire
a Bohm-Aharonov phase from the magnetic flux enclosed in the
region $\Gamma$.  At least in the adiabatic limit, the electron can not
distinguish the different sources of the two phases.\cite{kagome}
  The electron would
acquire the same total phase in the absence of
 the spin texture if instead an additional amount of flux
\begin{equation}
\Delta\Phi = \frac{\Omega}{4\pi} \Phi_0,
\label{eq:eff-flux}
\end{equation}
(where $\Phi_0$ is the flux quantum) were added to the region $\Gamma$.
We know however that adding flux to a region in a system with a finite
Hall conductivity
changes the total charge in that region.\cite{foot:adiabatic}
  To see this, let
$\Phi(t)$ be the time-dependent flux inside $\Gamma$.  Then the electric
field along the perimeter obeys (from Faraday's Law)
\begin{equation}
\oint_{\partial\Gamma} {\bf E}\cdot d{\bf r} = -\frac{1}{c}
\frac{d\Phi}{dt}.
\end{equation}
Because of the Hall conductivity (and the fact that $\sigma_{xx}=0$),
the field at the  perimeter induces a
current obeying
\begin{equation}
{\hat z}\cdot {\bf J}\times d{\bf r} = \sigma_{xy}\, {\bf E}\cdot d{\bf r}.
\end{equation}
Integrating this expression along the boundary and using
 the continuity equation, we have that the total charge inside $\Gamma$
obeys
\begin{equation}
\frac{dQ}{dt} = +\frac{\sigma_{xy}}{c} \frac{d\Phi}{dt},
\end{equation}
or
\begin{equation}
\Delta Q = - e\nu \frac{\Delta\Phi}{\Phi_0},
\label{eq:DQ}
\end{equation}
where we have used the fact that the Hall conductivity is quantized
(and is negative for ${\bf B} = |{\bf B}|{\bf {\hat z}}$):
\begin{equation}
\sigma_{xy} = -\nu e^2/h .
\end{equation}
 Thus $\nu$ electrons flow into the region
for each quantum of flux added to the region. This makes sense when we
recall
that there is one state in each Landau level per quantum of flux
penetrating the sample.

\null From Eq.(\ref{eq:eff-flux}) we see that the spin texture thus induces
an extra charge of
\begin{equation}
\Delta Q = - e\nu \frac{\Omega}{4\pi}.
\end{equation}
The solid angle $\Omega$ is of course
a functional of the spin texture in the region $\Gamma$.
For simplicity of analysis of this functional let us consider
making up $\Gamma$ out of a set of infinitesimal square loop circuits
in real space of the form
\begin{equation}
(x,y)
\longrightarrow
(x+dx,y)
\longrightarrow
(x+dx,y+dy)
\longrightarrow
(x,y+dy)
\longrightarrow
(x,y).
\end{equation}
The corresponding circuit in spin space illustrated in
Fig.[\ref{fig:stokes}] is,
\begin{equation}
{\bf m}(x,y)
\longrightarrow
{\bf m}(x+dx,y)
\longrightarrow
{\bf m}(x+dx,y+dy)
\longrightarrow
{\bf m}(x,y+dy)
\longrightarrow
{\bf m}(x,y).
\end{equation}
Approximating this circuit as a parallelogram as shown in
Fig.[\ref{fig:stokes}],
the solid angle subtended is (to a sufficient approximation)
\begin{equation}
d\omega = \left[{\bf m}(x+dx,y) - {\bf m}(x,y)\right] \times
 \left[{\bf m}(x,y+dy) - {\bf m}(x,y)\right] \cdot {\bf m}(x,y).
\end{equation}
This may be rewritten in a suggestive form
which tells us the curl of the Berry `connection:'\cite{berry}
\begin{equation}
d\omega = \frac{1}{2}\epsilon_{\mu\nu}\, {\bf m} \cdot
\partial_\mu {\bf m}\times  \partial_\nu {\bf m} \,\,dx\,dy.
\end{equation}
We can now add up all the infinitesimal contributions to obtain
\begin{equation}
\Omega = \int_\Gamma dx\,dy \,\,
 \frac{1}{2}\epsilon_{\mu\nu}\, {\bf m} \cdot
\partial_\mu {\bf m}\times  \partial_\nu {\bf m},
\end{equation}
which yields a total charge of
\begin{equation}
\Delta Q = -
\frac{e\nu}{8\pi}
 \int_\Gamma dx\,dy \,\,
\epsilon_{\mu\nu}\, {\bf m} \cdot
\partial_\mu {\bf m}\times  \partial_\nu {\bf m},
\end{equation}
or a local charge density deviation of
\begin{equation}
\delta \rho = -
\frac{e\nu}{8\pi}
\epsilon_{\mu\nu} \,{\bf m} \cdot
\partial_\mu {\bf m}\times  \partial_\nu {\bf m}.
\label{eq:charge-density}
\end{equation}
The expression on the right hand side of Eq.(\ref{eq:charge-density})
is simply the Pontryagin topological charge density of the spin texture.
Its integral over all space is an integer and is a topologically invariant
winding number known
as the Pontryagin index.  The spin textures which have non-zero Pontryagin
index are the `skyrmion' configurations illustrated in
Fig.[\ref{fig:spin-flip}b].  A microscopic variational wave function for
these spin textures was discussed in Section~\ref{sec:broken-symm}.

The charge density in Eq.(\ref{eq:charge-density})
can be viewed as the time-like component
of a conserved (divergenceless) topological `three-current'
which results in the following beautiful formula
\begin{equation}
j^\alpha = -\frac{\nu}{8\pi}\> \epsilon^{\alpha\beta\gamma}\>
\epsilon_{abc}\>
{m^a}({\bf r})
\partial_\beta {m^b}({\bf r})
\partial_\gamma {m^c}({\bf r}).
\label{eq:3.240a}
\end{equation}
Using the fact that $\bf m$ is a unit vector, it is straightforward to
verify that $\partial_\mu j^\mu = 0$.
We note that the fact that the
expression for the topological current is not parity invariant
is a direct reflection of the lack of parity symmetry in the presence of
the external magnetic field.

The  mechanism we have seen here that associates charge with flux is the reason
that quantum Hall fluids are described by Chern-Simons
theories,\cite{SMGchap10,zhang_cs,lz,lopez,kaneandlee}
and is the same mechanism which causes Laughlin
quasiparticles (which are topological vortices)
to carry quantized fractional charge proportional to the quantized value of
$\sigma_{xy}$.\cite{RBLchap7}

Having established that the electron charge density is proportional to
the topological density of the spin order parameter field, we must now
return to our Lagrangian to see what modifications this implies.  We have
already taken into account the long-range Coulomb force but it led only
to the local spin gradient term whose coefficient is the
spin wave stiffness.  There are however additional effects of the charge
(topological) density fluctuations which we must take into account
\begin{eqnarray}
{\cal L} &=&  \frac{\nu}{4\pi\ell^2}\int d^2r
 \frac{\partial {\bf m}}{\partial t} \cdot {\bf A}[{\bf m}]
-  \frac{1}{2} \rho_{\rm s}
\int d^2r [\nabla {m}^\mu({\bf r})]\cdot[\nabla
{m}^\mu({\bf r})]\nonumber\\
&+& \sum_{\bf q} V({\bf -q}) \delta\rho_{\bf q}
+ \frac{1}{2} \sum_{\bf q}  \frac{2\pi}{\epsilon q}
\delta\rho_{\bf -q} \delta\rho_{\bf q}.
\label{eq:charge-terms}
\end{eqnarray}
Here $\delta\rho_{\bf q}$ is the Fourier transform of the
charge density
in Eq.(\ref{eq:charge-density}).  Note that it is second order in spin
gradients.
The first of the new terms in Eq.(\ref{eq:charge-terms})
represents the coupling of the charge
fluctuations to the external and random disorder potentials $V(-{\bf q})$
and is
second order in spin gradients.  The second new term represents the
mutual interaction of the charge fluctuations via the Coulomb potential.
Note that this is fourth-order in gradients (and so is not a duplicate of
the $\rho_{\rm s}$ term which also comes from the Coulomb interaction).
  In general there will be additional fourth-order
terms  allowed by symmetry,
but we do not bother to write them down since they will not
have the divergent Coulomb interaction coefficient $2\pi/\epsilon q$ which makes the
term we have kept effectively third order in $q$.

We can immediately conclude several interesting
 things from the rather peculiar nature of
our itinerant ferromagnet.  First, unlike the case of a regular ferromagnet,
a scalar potential can induce the formation of charged
skyrmions in the ground state.  Thus sufficiently strong disorder
would have the effect of greatly
reducing the net spin polarization of the ground state,  something
which should be directly observable experimentally.

Secondly, we note that (at the classical level) the energy of a skyrmion
due to the gradient term is scale invariant
\begin{equation}
E = \frac{1}{2} \rho_{\rm s}
\int d^2r [\nabla {m}^\mu({\bf r})]
\cdot[\nabla {m}^\mu({\bf r})]
= \frac{E_0}{4},
\end{equation}
because we have two spatial integrations and two derivatives. [The quantity
$E_0$ is defined in Eq.(\ref{eq:E_0}).]
Now however the Coulomb self-energy will want to expand the size of the
skyrmion.  In real spin systems this effect competes against the small
but (usually) non-zero Zeeman coupling which wants to minimize the number
 of flipped spins.  This competition has been studied in some
detail and appears to be essential to explain the experiments
of Barrett et al.\cite{sondhi,fertig,barrett_prl,barrett_science}

\section{Interlayer Coherence in Double Layer Systems}
\label{sec:doublelayers}

The details of the double layer experiments of Murphy et al.\cite{murphyetal}
are described in the Chapter by J. P. Eisenstein.  Here we briefly
introduce the main ideas.  Double layer quantum Hall systems (and wide
single well systems\cite{mansour}) exhibit a variety of non-trivial
collective states
at different filling factors.
Here we will focus on the case of total filling factor
$\nu=1$ (that is, 1/2 in each layer) which is most closely analogous to the
fully ferromagnetic broken symmetry state for $\nu=1$
with `real' spins that we have discussed in the previous sections.
There are many other interesting states which we do not have room to
discuss here.
One example, is the state at total filling factor $\nu=1/2$ (that is, 1/4 in
each layer) which is  believed to be described by Halperin's $\{3,3,1\}$
wave function.\cite{halperinz1,macdsurface,chakrabortybook,nu1/2refs,chakra87} 
 This state
is more nearly like a gapped spin liquid state, although, as we have
already seen, it does not satisfy the Fock cyclic condition and so is not a true
spin singlet.

The schematic energy level diagram for the growth direction
degree-of-freedom in the double-layer system is shown
in Fig.[\ref{fig:double-well}].  For simplicity we assume that electrons
can only occupy the lowest electric subband in each quantum well.
If the barrier between the wells is not too strong,
tunneling from one side to the other is allowed. The lowest energy
eigenstates split into symmetric and antisymmetric combinations separated
by an energy gap $\Delta_{\rm SAS}$ which can, depending on the sample,
vary from essentially zero to hundreds of Kelvins.  The splitting
can therefore be much less than or even greater
 than the interlayer interaction
energy scale, $E_{\rm c} \equiv e^2/(\epsilon d)$.

The analogy with the spin systems studied in the previous sections is that
within the approximations just mentioned, the electrons
in a two-layer system have a double-valued internal quantum number---namely
the layer index.  The tunnel splitting in the double well
plays the role of the Zeeman splitting for spins.

In addition to double quantum wells, there are wide single
quantum wells in which the two lowest electric subband states are strongly
mixed by Coulomb interactions.\cite{mansour}  These systems
exhibit very similar physics, and can also be approximately modeled as
a double layer system.

Throughout our discussion we will
assume that the `real' spins
are aligned and their dynamics frozen out by the small Zeeman energy.
This is not necessarily a good approximation in experimentally relevant
cases but greatly simplifies matters.  Dynamics of `real' spins
in double layers is currently a topic of investigation.\cite{aoki_dbl,andersK}

\subsection{Experimental Indications of Interlayer Phase Coherence}
\label{subsec:expt_background}

Here we will very briefly  review the main experimental indications that
double well and wide single well  systems at $\nu=1$
can show coherent pseudospin
phase order over long length scales
and exhibit excitations which are highly collective in
nature.

When the layers are widely separated, there will be no correlations between
them and we expect no dissipationless quantum Hall state\cite{nuhalf}
since each layer has $\nu = 1/2$.  For smaller separations, it 
was predicted theoretically\cite{chakra87,yosh89,fertigold} and subsequently
observed experimentally that there is an excitation gap and a quantized
Hall plateau.\cite{greg,mansour,murphyetal}  The resulting phase diagram
 is shown in
Fig.[\ref{fig:qhe-noqhe}] and discussed in more detail in the Chapter in
this volume by J. P. Eisenstein.
The existence of a gap has either a trivial or a
highly non-trivial explanation, depending on the ratio
$\Delta_{\rm SAS}/ E_{\rm c}$.   For large $\Delta_{\rm SAS}$
the electrons tunnel back and forth so rapidly that it is as if there is
only a single quantum well.  The tunnel splitting $\Delta_{\rm SAS}$
is then analogous to the electric subband splitting in a (wide) single
well.  All symmetric states are occupied and all antisymmetric states are
empty and we simply have the ordinary $\nu = 1$ integer Hall effect.
Correlations are irrelevant in this limit and the excitation gap is close
to the single-particle gap $\Delta_{\rm SAS}$  (or $\hbar\omega_{\rm c}$,
whichever is smaller).
What is highly non-trivial about this system is the fact that the
$\nu = 1$ quantum Hall plateau survives even when $\Delta_{\rm SAS}
\ll E_{\rm c}$ (See Fig.[\ref{fig:qhe-noqhe}]).
  In this limit the excitation gap has clearly changed to
become highly collective in nature since the
observed\cite{mansour,murphyetal}
gap can be on the scale of 10-20 K even when $\Delta_{\rm SAS} \sim
1{\rm K}$.  Because of the spontaneous broken
symmetry,\cite{Wen-Zee_double,ezawa,ezawa_95,harfok,usPRL} 
the excitation gap actually
survives the limit $\Delta_{\rm SAS} \longrightarrow 0$!
This cross-over from single-particle to purely collective gap is
quite analogous to the result we discussed earlier that for spin polarized
single layers, the excitation gap survives the limit of zero Zeeman
splitting.  Hence, to borrow a delightful phrase from 
Sondhi {\it et al.}\cite{sondhi_phrase}, `$\nu=1$ is a fraction too.'

A second indication of the highly collective nature of the excitations
can be seen in the Arrhenius plots showing thermally activated
dissipation.\cite{murphyetal,JPEchap}  The low temperature activation energy
$\Delta$ is, as already noted, much larger than $\Delta_{\rm SAS}$.  If
$\Delta$ were nevertheless somehow a single-particle gap, one would expect
the Arrhenius law to be valid up to temperatures of order $\Delta$.
Instead one observes a fairly sharp leveling off in the dissipation as the
temperature increases past values as low as $\sim 0.1 \Delta$.  This is
consistent with the notion of a thermally induced collapse of the order
that had been producing the collective gap.  This behavior is very similar
to that seen in `real' spins.\cite{barrett_prl,barrett_science}

The third significant feature of the experimental data pointing to a
highly-ordered collective state is the strong response of the the system
to relatively weak magnetic fields $B_\parallel$ applied in the plane of
the 2D electron gases.  Within a model that neglects higher electric
subbands, we can treat the electron gases as strictly two-dimensional.
This is important since $B_\parallel$ can then
affect the system only if there
are processes that carry electrons around closed loops containing flux.  A
prototypical such process is illustrated in Fig.[\ref{fig:parallelB}].
  An electron
tunnels from one layer to the other at point A, and travels to point B.
Then it (or another indistinguishable electron) tunnels back and returns
to the starting point.  The parallel field contributes to the quantum
amplitude for this process (in the 2D gas limit) a gauge-invariant
Aharonov-Bohm phase factor $\exp\left(2\pi i \Phi/\Phi_0\right)$ where
$\Phi$ is the enclosed flux and $\Phi_0$ is the quantum of flux.

Such loop paths evidently contribute significantly to correlations in the system
since the activation energy gap is observed to decrease very rapidly with
$B_\parallel$, falling by factors of order two or more until a critical field,
$B^*_\parallel \sim 0.8{\rm T}$, is reached at which the gap essentially ceases
changing.\cite{murphyetal}  To understand how remarkably small
$B^*_\parallel$ is, consider the following.  We can define a length
$L_\parallel$ from the size of the loop needed to enclose one quantum of flux:
$L_\parallel  B^*_\parallel d = \Phi_0$. ($L_\parallel [\hbox{\AA}]
= 4.137 \times 10^5
/ d [\hbox{\AA}]
 B^*_\parallel [\rm T]$.) For $B^*_\parallel = 0.8{\rm T}$ and $d = 210
\hbox{\AA}$,
$L_\parallel = 2460 \hbox{\AA} $
which is approximately six times the spacing
between electrons in a given layer and more than twenty
times larger than the quantized cyclotron orbit radius
$\ell \equiv (\hbar c / e B_\perp)^{1/2}$ within an individual layer.
Significant drops in the excitation gap are already seen at fields of 0.1T
implying enormous phase coherent correlation lengths must exist.  Again this
shows the highly collective nature of the ordering in this system.

The fourth indication about the nature of the coherent ordering is the fact
that the gapped quantum Hall state at $\nu=1$ can survive a finite amount
of imbalance in the layer charge densities.\cite{JPEchap}
  Charge imbalance can be
controlled by applying a gate voltage which increases the
equilibrium charge density in one layer and decreases it in the other.
In some cases, such as the gapped state at $\nu=1/2$, layer imbalance
immediately destroys the quantum Hall state.  For $\nu=1$ it does not.
This is a strong hint about the different
natures of the ordering in the two cases.  We defer discussion of
what this hint means to subsection \ref{subsec:superfluid_dynamics}.

\subsection{Effective action for double layer systems}
\label{subsec:easyplane}

Having established the correct form of the effective
low-energy Lagrangian for the SU(2) invariant case, let us now turn
to the U(1) symmetric case in double layer systems.  The spin analogy
is relatively straightforward, but can be confusing until one gets used to it.
The idea is simply that each electron can be in either the upper layer
or the lower layer and we refer to these two states as pseudospin up and
down respectively.  For the moment we shall
 assume the layers are identical and also
neglect the possibility of tunneling between the two layers.
The confusing
point is that quantum mechanics nevertheless forces us
to consider the possibility that an electron can be in a coherent
superposition of the two pseudospin states so that its layer index is
uncertain.

Portions of
our discussion here and in the following subsections follows that of
Ref.[\citen{uslongI}].  Further technical details can be found therein.
The formal  mapping that we use to define the pseudospin density operators is
the following.
The $z$ component of the pseudospin density represents the local charge density
difference between the layers
\begin{equation}
S^z ({\bf r}) = \frac{1}{2}\left[
\psi_\uparrow^\dagger ({\bf r}) \psi^{\phantom{\dagger}}_\uparrow ({\bf r}) -
\psi_\downarrow^\dagger ({\bf r})
\psi^{\phantom{\dagger}}_\downarrow ({\bf r})\right].
\end{equation}
The $x$ and $y$ components of the pseudospin density are off-diagonal
and can be combined to form tunneling operators
\begin{equation}
S^{+}({\bf r})
= \left[S^{-}({\bf r})\right]^\dagger
= \frac{1}{2}
\psi_\uparrow^\dagger ({\bf r})
\psi^{\phantom{\dagger}}_\downarrow ({\bf r}).
\end{equation}
If for instance, tunneling is present then the Hamiltonian contains a term
\begin{equation}
T = -t \int d^2r\,\left[S^+({\bf r}) +
S^-({\bf r})\right]
= -2t \int d^2r\,S^x({\bf r}).
\end{equation}

We know that ferromagnetism in real spin systems is
not the result of truly spin-dependent forces but rather a byproduct of
Coulomb
exchange forces. Hence the effective spin Hamiltonian must contain
only spin scalars such as ${\bf S}_i\cdot {\bf S}_j$.  Here however the
Coulomb forces are explicitly pseudospin dependent since the intra-
and inter-layer Coulomb interactions are not identical.
Following our previous discussion\cite{macd-expt} we define
\begin{equation}
V^0_k \equiv \frac{1}{2}\; (V^A_k+V^E_k),
\end{equation}
\begin{equation}
V^z_k \equiv \frac{1}{2}\; (V^A_k-V^E_k),
\end{equation}
where $V^A_k$ is the Fourier transform with respect to the
planar coordinate of the
interaction potential between a pair of
electrons in the same layer and $V^E_k$ is the Fourier
transform of the interaction potential between a pair of
electrons in opposite layers.   If we neglect the finite
thickness\cite{macd-expt} of the layers, $V^A_q=2 \pi e^2 /(\epsilon q)$
and $V^E_q = \exp (-qd) V^A_q$,
where $d$ is the layer separation.
The interaction Hamiltonian can then be separated into
a pseudospin-independent part with interaction $V^0$ and
a pseudospin-dependent part.  The pseudospin dependent term in
the Hamiltonian is
\begin{equation}
\overline V_{\rm sb}
=2\sum_{k}{V_{k}^{z}\overline S_{k}^{z}\overline S_{-k}^{z}}.
\label{eq:Vsymm-break}
\end{equation}
Here ${\bar {\bf S}_{\bf k}}$ is the Fourier transform of the spin density
at wave vector ${\bf k}$ and the overbar indicates projection onto the
lowest Landau level.\cite{uslongI}
Since $V^A_k >V^E_k$, $V^z_k$ is positive and this term produces an
easy-plane, as opposed to Ising, pseudospin
anisotropy.  That is, this term prefers
for the spin to lie in the XY plane.  If the spin orientation
moves up out of the XY
plane so that $\langle S^z \rangle \ne 0$, then the energy increases.  We
can view this energy cost as simply the charging energy of the capacitor
formed by the two layers since the pseudospin component
$S^z$ measures the charge difference between the two layers.

The pseudospin symmetry of
the Hamiltonian is reduced from $SU(2)$ to $U(1)$ by this term.
In addition, this term increases the quantum fluctuations in the
system since it does not commute with the order parameter
\begin{equation}
[\overline V_{\rm sb}, S^\mu] \ne 0,
\end{equation}
where $\mu = x,y$.  Thus total spin is no longer a sharp quantum number.
However for small layer separations, we expect the quantum fluctuations to be
small.  At very large layer separations we expect the quantum fluctuations
to become dominant and produce the
 disordering phase transition which uncouples
the two layers.

In the absence of the symmetry breaking term ($d=0$), we know the exact quantum
ground state for $\nu=1$ since it is simply a fully occupied
pseudospin-polarized Landau level.
This state is $2S+1= N+1$ fold degenerate
(since there is no charging energy in this
unphysical limit).
As a first approximation for finite $d$,
we assume that the eigenstates remain exactly the
same, but the charging energy lifts the degeneracy among these states by
favoring the ones  with $S^z\sim 0$.  That is, we assume that the spin vector is
still fully polarized and non-fluctuating, but it now lies in the XY plane.

We argue that the
form of the energy-functional we derive must remain valid
even when quantum fluctuations due to the pseudospin-dependent
terms in the Hamiltonian are present.  However, the coefficients
which appear in the energy-functional will be altered by
quantum fluctuations and the explicit expressions we derive
below are accurate only when the pseudospin-dependent interactions
are weak, {\it i.e.}, only when the layers are close together.
Estimates have been
obtained for the quantum fluctuation corrections to
these coefficients from finite-size
exact diagonalization and many-body perturbation theory
calculations.\cite{uslongI,uslongII}

To better understand what we are assuming,  consider the specific example
of a state which has $S=N/2$ and has its spin oriented in the
${\bf\hat x}$ direction so that it is an eigenstate of total $S^x$
\begin{equation}
\big|\Psi\big\rangle =
\prod_X \frac{1}{\sqrt{2}}
 \left(c^\dagger_{X\uparrow} + c^\dagger_{X\downarrow}\right)
\big|0\big\rangle.
\label{eq:variational}
\end{equation}
Here $X$ is a LLL orbital label and $|0\rangle$ is the electron vacuum.
It is clear from the construction that this state has every orbital
in the LLL filled so it has $\nu=1$.  However each electron is in
a linear combination of up and down pseudospin which puts it in
an eigenstate of $S^x$.  This particular combination is familiar because
it is the symmetric state which is the exact (non-interacting)
ground state in the presence of
tunneling between the layers.  In the absence of $V_{\rm sb}$ this would be
an exact eigenstate of the system (with or without tunneling) even in the
presence of the symmetric interaction term $V_0$ because it is
a filled Landau level.  This state has zero net
charge on the capacitor on the average ($\langle S^z\rangle=0$) and so
first-order perturbation theory in $V_{\rm SB}$ favors it for the ground
state.

The confusion that many people have at this point is over the fact that
this is a coherent state with uncertain $S^z$. It makes perfect sense
if the Hamiltonian includes tunneling that the charge in each layer will be
uncertain.  However we are arguing that this state is a good approximation
to the ground state even in the absence of tunneling because it has
favorable Coulomb energy.  That is, we are assuming that the system
spontaneously breaks the $U(1)$ symmetry associated with
conservation of layer charge 
difference\cite{Wen-Zee_double,ezawa,ezawa_95,fertigold}
and is acting much like a superfluid\cite{macd-expt} which breaks $U(1)$
symmetry.  We refer to this state as having spontaneous interlayer phase
coherence.  It is quite analogous to a BCS state which has an uncertain
total number of Cooper pairs.  In the absence of tunneling, the electron
energy cannot be sensitive to the actual relative phase between the two
layers and so any coherent state of the form
\begin{equation}
\big|\Psi\big\rangle =
\prod_X \frac{1}{\sqrt{2}} \left(c^\dagger_{X\uparrow} +
e^{i\varphi}c^\dagger_{X\downarrow}\right) \big|0\big\rangle
\label{eq:variationalpsi}
\end{equation}
will have equally good energy.  The global U(1)
phase $\varphi$ simply determines
the orientation of the total spin in the XY plane
\begin{equation}
\langle\Psi|S^+|\Psi\rangle \equiv
\langle\Psi| S^x + i S^y|\Psi \rangle = \frac{N}{2}e^{i\varphi}.
\label{eq:orderparam}
\end{equation}
This is exactly analogous to the situation in a BCS superconductor
where the energy is independent of the phase $\varphi$ which determines
the coherence between states of different numbers of Cooper pairs
\begin{equation}
\left\langle \psi^\dagger_\uparrow\psi^\dagger_\downarrow\right\rangle
\sim e^{i\varphi}.
\end{equation}
The complex order parameter in a double layer system given
by Eq.(\ref{eq:orderparam})
has an amplitude and a phase like that of a superconductor, but in many ways is
more reminiscent of an excitonic insulator since it is charge neutral
(i.e., it contains $\psi^\dagger_\uparrow \psi_\downarrow$
not $\psi^\dagger_\uparrow\psi^\dagger_\downarrow$)
and physically
represents a particle in one layer bound to a hole in the other layer
(but we do not know which layer contains the particle and which
contains the hole).  Some years ago, Datta\cite{datta} considered similar
states in double layer systems in zero B field.

At first, it may seem counter-intuitive
that, in the absence of tunneling, the system energy
could depend on the relative phase $\varphi$ for finding the particle in
the upper or lower layer.  Indeed the energy is unchanged when a constant
is added to $\varphi$.  However the energy does depend on gradients of
$\varphi$.  The `spin stiffness' $\rho_s$ is non-zero because of the loss
of exchange energy which occurs when $\varphi$ varies with position.
Imagine that two particles approach each other.  They are in a linear
superposition of states in
each of the layers (even though there is no tunneling!).
If these superpositions
 are characterized by the same phase, then the wave function is
symmetric under pseudospin exchange and so the spatial wave function
is antisymmetric and must vanish as the particles approach each other.
This lowers the Coulomb energy.  If a phase gradient exists, then there is
a greater amplitude for the particles to be near each other and hence the
energy is higher.  This loss of exchange energy is the source of the finite
spin stiffness and is what causes the system to spontaneously `magnetize'.
We may describe this by a term in the effective action of the form
\begin{equation}
E = \frac{1}{2} \rho_s \int d^2 r\, |\nabla\varphi|^2.
\label{eq:2dxy}
\end{equation}
We see immediately that at finite temperatures this system will be
described by a classical XY model and will undergo a Kosterlitz-Thouless
phase transition.  We will cover this in more detail further below.

\subsection{Superfluid Dynamics}
\label{subsec:superfluid_dynamics}

We consider now the dynamics of a two layer system with easy-plane
pseudospin
anisotropy but no tunneling.
Let us back up and note that implicit in Eq.(\ref{eq:2dxy})
 is the assumption that the spin lies exactly in the XY plane.
To be more complete we should derive the effective Lagrangian in the
presence of the easy plane anisotropy.  On general symmetry grounds this
must be of the form (neglecting higher order gradients needed to
describe charge fluctuations)
\begin{equation}
{\cal L} =  \frac{\nu}{4\pi\ell^2}\int d^2r
 \frac{\partial {\bf m}}{\partial t} \cdot {\bf A}[{\bf m}]
-  \int d^2r
\left\{\beta (m^z)^2 +
\frac{\rho_{\rm A}}{2}|\nabla m^z|^2
+ \frac{\rho_{\rm E}}{2}[|\nabla m^x|^2+|\nabla m^y|^2]\right\}.
\label{eq:xylagrangian}
\end{equation}
Within the Hartree-Fock approximation which is presumably valid
only for small layer separations, the coefficients $\beta,
\rho_{\rm A}$, and $\rho_{\rm E}$
can be evaluated in terms of the density-density
correlation function of the ground state.\cite{uslongI}
(The gradient expansion in $m^z$ is not strictly valid since
it turns out\cite{uslongI} that
the long range of the Coulomb interaction leads to a non-local
$m^z m^z$ interaction term not included above.  This term vanishes
in the SU(2) invariant case but here is 
more important than the $|\nabla m^z|^2$ term and less important than
the $\beta |m^z|^2$ term at long wavelengths.  
We retain the $|\nabla m^z |^2$ term in the following expressions only 
to remind us of its importance in the SU(2) invariant limit.) 

Since $m^z$ is massive, the equations of motion derived from this
Lagrangian lead to a linear rather than quadratic collective mode
dispersion\cite{fertigold,macd-expt,uslongI,Wen-Zee_double,ezawa,ezawa_95,
jasonho}
like that of the Goldstone mode in a superfluid or an antiferromagnet.
We take the pseudospin of the system to be polarized in the
${\bf \hat x}$ direction and consider the linear response to a time--
and space--dependent pseudospin
`Zeeman' field in the ${\bf\hat y} - {\bf\hat z}$ plane.
Using the equations of motion determined from the Lagrangian in
Eq.(\ref{eq:xylagrangian}) and
Fourier transforming with respect to both time and space we find that
\begin{equation}
\left(\begin{array}{cc}
- i \omega & - { 4 \pi  \over \nu } ( 2 \beta + q^2 \rho_A) \\
{ 4 \pi \over \nu } ( q^2 \rho_s) & - i \omega \\
\end{array} \right)\;
\left(\begin{array}{c} m_y \\ m_z \\ \end{array} \right)
=
\left(\begin{array}{c} - h_z \\ h_y \\ \end{array} \right),
\label{eq:resfunc}
\end{equation}
where $h_y$ and $h_z$ are the Fourier coefficients of the
pseudospin magnetic field at frequency $\omega$ and wavevector
$q$.  Physically $h_z$ corresponds to a time-- and space--dependent
bias potential between the two wells, while
$h_y$ could arise from a space-- and time--dependent interlayer
tunneling amplitude.  We see immediately that the response has a
singularity
at the collective mode frequency
\begin{equation}
\omega_q^2   = \left({ 4 \pi \over \nu }\right)^2
\left[2 \beta + q^2 \rho_A\right]  q^2 \rho_s.
\label{eq:colmode}
\end{equation}
For the $d/ \ell = 0$ case, $\beta = 0$, $\rho_A = \rho_s =\rho_s^0$,
and the collective mode frequency reduces to the result
obtained previously for the spin-wave collective mode of isotropic
ferromagnets [$\omega_q  = 4 \pi q^2 \rho_s^0 / \nu $].
The collective mode corresponds to a spin-precession whose
ellipticity increases as the long-wavelength limit is
approached.  The presence of the mass term ($\beta \ne 0$)
changes the collective
mode dispersion at long wavelengths from quadratic to
linear.\cite{macd-expt}  In the limit of small $q$
\begin{equation}
\omega_q={ 4\pi \over \nu} \sqrt{2\beta\, \rho_s }\,\, q.
\end{equation}

Thus we see that the system is acting very much like a bosonic
superfluid of weakly repulsively interacting particles which has a
Goldstone mode (if and only if there is non-zero
repulsion).\cite{macd-expt,Wen-Zee_double,ezawa,%
ezawa_95,usPRL,uslongI,jasonho,du}
However we must again emphasize that we have a charge neutral order
parameter.  The singular response occurs for electric fields of opposite
signs in the two layers.  Another way to say this is that the `charge'
conjugate to the U(1) phase field $\varphi$ is $S^z$.  The current
associated with phase gradients,
\begin{equation}
J_{zz} = \frac{2\rho_s}{\hbar} \nabla \varphi,
\label{eq:superflow}
\end{equation}
is the flow of pseudospin density which is the difference between the
electrical currents in the two layers.\cite{uslongI,jasonho,ezawa,ezawa_95}
  This simply reflects the fact
that the object which condenses is a neutral excitonic composite of a
particle and hole.

These analogies with superfluidity naturally raise the question of
analogs of other phenomenon known to occur in superfluids and
superconductors.  
It is believed that there is no analog of the Meissner effect in these
systems.\cite{usPRL,uslongI,ezawa_95}
However, it has been suggested that
Josephson or `Josephson like'  effects should occur 
in the tunneling transport between
the two layers.\cite{Wen-Zee_double,ezawa,ezawa_95}  In this picture one views
each layer as analogous to one side of a Josephson junction.
This appealing analogy appears to imply measureable consequences only
in the case where superconducting contacts are made to 
each two-dimensional electron layer;
the distinction between the system and an ordinary Josephson junction system
then seems to be artificial.  More interesting, in our view, is 
the recent `sideways tunneling' proposal of Wen and
Zee\cite{sideways_tunneling} which appears to provide
a quite precise analog of the Josephson effect, but in this case for
pseudospin superfluidity.
Here one views the {\em pair} of layers as
constituting a single superfluid on
one side of a Josephson junction.  A separate pair of layers
is imagined to be coupled to the first by weak `sideways' tunneling.
In each layer, there is a phase variable $\varphi$ and a pseduospin
supercurrent given by Eq.(\ref{eq:superflow}).  When this current
reaches the junction, it can coherently
jump across the weak link and continue onwards.  Physically this coherent
tunneling of pseudospin involves an electron tunneling one way between
the upper layers and another electron tunneling in the {\em opposite}
direction between the lower layers.  It should be possible in principle to
see AC Josephson oscillations of the current 
at a frequency of $2eV/\hbar$ where $V$ an appropriately applied bias
voltage (of opposite sign in the two layers).  It is unlikely however
that a sample with no tunneling within each pair of layers, 
but finite tunneling across the junction, can ever be produced.  
One could imagine
using a parallel B field to effectively shut off the tunneling
within each pair, but even weak disorder will pin the discommensurations
and probably ruin the effect.

As an aside to the question of spin channel superfluidity,
we point out that double layer systems are ideal for
probing electron-electron interactions via mutual drag effects in
transport at zero field.\cite{drag} 
More details on some of the mutual transport effects associated 
with spontaneous interlayer phase coherence are presented in
Ref.[\citen{uslongI}].  In addition, it has been predicted that 
for double-layer fractional quantum Hall states without broken 
symmetries, but with 
inter-layer correlations of the type described by the Halperin 
$\{m,m',n\}$ wavefunctions,
anomalously large (and quantized!) mutual drag 
effects will occur.\cite{renndbl,du}

We are now in a position to return to a discussion of the fourth
experimental indication of interlayer phase coherence that was briefly
alluded to at the end of subsection~\ref{subsec:expt_background},
namely the fact that unlike the $\nu=1/2$ state,
the $\nu=1$ state survives a finite amount of charge imbalance in the two
layers.  The rigidity of the $\nu=1/2$ state  can be understood
macroscopically from the Chern Simons effective
field theory point of view,\cite{chern-simonsnu=1/2refs,Halperin_Newport}
or microscopically from the fact that the Halperin $\{3,3,1\}$ state is
believed to give a good description of the state.
The plasma analogy shows
that this state has a sharply defined total filling $\nu=1/2$ and
that the density in each layer must be exactly equal (see
Section~\ref{sec:frac-charge}).  Another way to
say this is that the $\{3,3,1\}$ state is almost a singlet spin liquid
which must have $S^z=0$
(although strictly speaking, it is not an eigenstate of total spin).
That is to say, it is pseudospin incompressible.

In contrast to this, the $\nu=1$ state is pseudospin compressible.
This is clear from the existence of pseudospin superfluidity with
its gapless collective Goldstone mode.
The state is ferromagnetically ordered and there is a whole family
of states differing only in $S^z$ which would be degenerate were it not for
the charging energy.  The charging energy lifts this degeneracy
and picks out the $S^z=0$ state.  Application of a bias voltage to
unbalance the layers, simply picks a different state
having non-zero $S^z$ out of the manifold.

We can supplement this picture by thinking
macroscopically of the XY model describing the ordering.  Layer imbalance
corresponds to tilting the order parameter slightly up out of the XY
plane.  This changes no symmetries in the problem, and simply renormalizes
the spin stiffness slightly because the projection of the spin onto the XY
plane is reduced.  Thus the ordering which produces the charge gap
is weakened but not (immediately)
destroyed by layer imbalance.  One can also describe this
microscopically by modifying the variational wave function in
Eq.(\ref{eq:variationalpsi}) to the form
\begin{equation}
\big|\Psi\big\rangle =
\prod_X \frac{1}{\sqrt{1+\gamma^2}}\left(c^\dagger_{X\uparrow} +
\gamma e^{i\varphi}c^\dagger_{X\downarrow}\right) \big|0\big\rangle,
\label{eq:variationalpsi2}
\end{equation}
which obeys
\begin{equation}
\left\langle\Psi\left| S^z \right|\Psi\right\rangle = \frac{N}{2}\left(
1-\gamma^2\right).
\end{equation}

We close by noting that
a class of experiments that would be very useful in probing the
pseudospin superfluidity
would be inelastic (Raman) light scattering
to try to detect the gapless collective excitation mode.  To see this
at finite frequency requires finite
wavevectors which in turn would require some sort of grating coupler.
These are difficult to fabricate on sufficiently short length scales.
Here however,
it turns out that it is  possible to have the system act as
its own grating by tilting the applied magnetic field which will induce
a twisting of the order
parameter\cite{parallel_coll_modes,uslongII}  as we will discuss
in  Section~\ref{sec:parallelB}.
\subsection{Merons:  Charged Vortex Excitations}
\label{subsec:merons}

Continuing the superfluid analogy, we study vortex excitations in
this section and discuss the Kosterlitz-Thouless phase
transition induced by the unbinding of these topological
defects.\cite{Wen-Zee_double,ezawa,usPRL,uslongI}
The order parameter of the system in the presence of a vortex at the origin
has the following approximate form
\begin{equation}
{\bf m}=\left\{\,\pm\sqrt{1-[m^z(r)]^2} \cos\theta\, ,
\sqrt{1-[m^z(r)]^2} \sin\theta\, , m^z(r)\right\}.
\label{meronvariational}
\end{equation}
Here the $\pm$ refers to right and left handed vortices respectively, and
$\theta$ is the azimuthal angle made by the position vector ${\bf r}$.
At asymptotically large radii, $m^z$ vanishes to minimize the charging
energy.  However in the vortex core we must have $m^z \longrightarrow \pm
1$  (and $m^x, m^y \longrightarrow 0$)
to prevent a singularity in the gradient energy.  Thus there are four
flavors of topologically stable objects which we refer to as `merons' since
it turns out that they are essentially half skyrmions.  These are
illustrated in Fig.[\ref{fig:meron}].

The local topological charge density
calculated from $\delta\rho= -{1\over 8\pi}\epsilon_{ij}(\partial_{i}{\bf m}
\times\partial
_{j}{\bf m})\cdot{\bf m}$ can be expressed in the form
\begin{equation}
\delta\rho(r)=\frac {1}{4\pi r} \frac {d m^z}{d r},
\end{equation}
and the total charge is
\begin{equation}
Q=\int d^{2}r\, \delta\rho(r)=\frac {1}{2} \bigr[m^z(\infty)-m^z(0)\bigl].
\end{equation}
For a meron, the spin points up or down at the
core center and tilts away from the ${\bf\hat z}$ direction
as the distance from the core center increases.
At asymptotically large distances from the origin, the spins
point purely radially in the ${\bf\hat x} -  {\bf\hat y}$ plane.
Thus the topological charge is $\pm \frac {1}{2}$ depending on the
polarity of core spin.
The general result for the topological charge of the four meron
flavors may be summarized by the following formula:
\begin{equation}
Q=\frac {1}{2} \bigr[m^z(\infty)-m^z(0)\bigl]\, n_{\rm v},
\end{equation}
where $n_{\rm v}$ is the vortex winding number.
The formulae derived above
for the meron charge do not rely for their validity on the variational
ansatz assumed in Eq.(\ref{meronvariational}).  They are quite general and
follow from the fact that a meron topologically has half the spin winding of a
skyrmion.  The meron  charge of $\pm 1/2$ is a topological invariant and
implies that the electrical charge is $\pm \nu e/2$.

The fact that merons carry fractional charge $\pm \nu e/2$ can be deduced
from a Berry phase argument similar to the one used to find the skyrmion
charge.  We simply note that an electron moving at a large distance
around a meron will have its spin rotated through $2\pi$ in the 
$\hat x- \hat y$ plane
due to the vorticity.  We know that the Berry's phase for rotating a spin
one-half object in such a way is $\exp{(i 2\pi S)} = -1$.  Thus the meron
produces the same Berry's phase as half a flux quantum.
\null From Eq.(\ref{eq:DQ}) we then obtain $\Delta Q = \pm \nu e/2$.
[The ambiguity of the sign of the charge associated with half a flux
quantum can only be resolved by examining the behavior of $m^z$ in the
meron core.  It depends on whether the mid-gap state induced by the
topological defect (discussed below) is empty or occupied.]

It is instructive to  write down explicit
microscopic variational
wave functions for vortices (merons). We start with the simplest example:
a meron with vorticity +1 and charge $-{1\over 2}$ that has the smallest
possible core size:
\begin{equation}
\vert\Psi_{+1,-{1\over 2}}\rangle
=\prod_{m=0}^M\left({1\over \sqrt{2}}c^{\dagger}_{m\uparrow}
+{1\over \sqrt{2}}c^{\dagger}_{m+1\downarrow}\right)\vert 0\rangle.
\label{smglabel1}
\end{equation}
Here $\vert 0\rangle$ is the fermion vacuum,
$c^{\dagger}_{m\uparrow(\downarrow)}$
creates an electron in the upper (lower) layer in the angular momentum $m$
state in the LLL, and $M$ is the angular momentum quantum number
corresponding to the edge.
The vorticity is +1 because far away the spin wave function is essentially
\begin{equation}
\chi(\theta)={1\over \sqrt{2}}\left(\begin{array}{c}
e^{+i\theta}\\
1\end{array}\right),
\end{equation}
where $\theta$ is the polar angle.
The charge is $-{1\over 2}$ because we have created a hole in the center of the
lower layer (every state has occupancy $1/2$ except
$m=0\downarrow$ which is unoccupied).
Since the spin is pointing up at the center, this agrees with
the spin-charge relation derived earlier. From the spin-charge relation we
know we can change the sign of the charge of a meron by changing the direction
of spins in the core region without changing the vorticity. This can be seen
explicitly from the wave function:
\begin{equation}
\vert\Psi_{+1,+{1\over 2}}\rangle
=c^{\dagger}_{0\downarrow}\prod_{m=0}^M
({1\over \sqrt{2}}c^{\dagger}_{m\uparrow}
+{1\over \sqrt{2}}c^{\dagger}_{m+1\downarrow})\vert 0\rangle.
\label{smglabel2}
\end{equation}
This state has charge $+{1\over 2}$ because we have put in an electron in the
$m=0$ state in the lower layer.
Obviously what we did
(in terms of the spin texture language)
is to flip the spins in
the core region to the down direction without changing the vorticity of the
meron at long distances.  In this construction, one sees that in a sense,
the merons are like fractionally charged `mid-gap states' in polyacetylene
which can be empty or occupied.\cite{schrieffer}

A meron with vorticity -1 is readily obtained by simply interchanging the
labels $m$ and $m+1$ in  the subscripts in Eq.(\ref{smglabel1}) and
Eq.(\ref{smglabel2}).  Invariance under pseudospin reversal
guarantees the equality of the
energies of corresponding $\pm$ vorticity merons.  However the two
charge states for a given vorticity are not necessarily degenerate, just
as Laughlin quasiholes and quasiparticles are in general non-degenerate.
[For the special case $\nu=1$ particle-hole symmetry guarantees
degeneracy.]

It is instructive to attempt to find
a variational wave function for a pair of
merons to show that a meron is half a skyrmion.  Consider the situation
where we have a pair of merons of opposite vorticity but the same charge,
located at points ${\bar Z}_1$ and ${\bar Z}_2$.
To achieve this one meron
must have $m^z=+1$ in its core and the other must have
$m^z=-1$ in its core.  The following wave function seems to do the job:
\begin{equation}
\psi_\lambda = \prod_j \frac{1}{\sqrt{2}}
 \left(\begin{array}{c} e^{i\varphi}(Z_j-{\bar Z}_1) \\
 (Z_j - {\bar Z}_2) \\ \end{array}
\right)_j \Phi_{mmm}
\label{eq:meron-pair}
\end{equation}
where $\Phi_{mmm}$ is defined in Eq.(\ref{eq:halperin-multi-psi}), 
$\varphi$ is an arbitrary constant phase angle,
and $()_j$ refers to the spinor for the $j$th particle.
Note that at very large distances away from ${\bar Z}_1$ and ${\bar Z}_2$
the spinor for each particle asymptotically approaches
\begin{equation}
Z_j\frac{1}{\sqrt{2}} \left(\begin{array}{c} e^{i\varphi}\\1\\
 \end{array}\right),
\end{equation}
and so corresponds to a fixed spin orientation in the xy plane
at an angle $\varphi$ from the x axis.  Hence the net vortex content
of the pair of excitations is zero.  It is clear
from the construction that the spin orientation
is purely up for an electron located at ${\bar Z}_2$ and
purely down at ${\bar Z}_1$.
Furthermore the net
charge must be $+\nu e$ since asymptotically the factor of $Z_j$ is
the same one as for the Laughlin quasihole in a spin polarized state.
By symmetry, it seems that
there must be half that amount, $+\nu e/2$ associated with
each of the two objects located at ${\bar Z}_1$ and ${\bar Z}_2$.
(It turns out however, that the charge is not split in half and localized
near ${\bar Z}_1$ and ${\bar Z}_2$ as we would really want for a meron pair.)

We can easily show that this particular wave function actually
is just another way to represent
a skyrmion.  Let $Z_1=\lambda$ and $Z_2=-\lambda$ and, for simplicity,
assume that the spins are asymptotically oriented in the
${\bf \hat x}$ direction
so that $\varphi=0$.  Now perform a global rotation of all the spins
about the ${\bf\hat y}$ axis by an angle of $-\pi/2$. Using
\begin{equation}
\exp\left\{i\frac{\pi}{4}\sigma^y\right\}
  \frac{1}{\sqrt{2}}\left(\begin{array}{c} (Z_j-\lambda) \\
 (Z_j +\lambda)  \\ \end{array}\right)
=
 \left(\begin{array}{c} Z_j \\
 \lambda  \\ \end{array}
\right),
\end{equation}
we see that we have recovered the variational skyrmion wave function
studied previously in Eq.(\ref{eq:skyrmicro}).  The form 
in Eq.(\ref{eq:meron-pair}) is the
appropriate one in the U(1) symmetric case since it keeps the spins
primarily in the XY plane.
We expect that a pair of merons can be deformed into
a skyrmion and each meron can be properly viewed as carrying
 half the topological and electrical charge of a skyrmion.
 However there must exist variational wave functions which are better
 for the U(1) case than
 the one discussed here in the sense that the present one has a smooth
 charge distribution centered in between  ${\bar Z}_1$ and ${\bar Z}_2$
 rather than being more closely associated with the two points defining the
 meron positions.

Finally we note that merons carry fractional statistics 1/4; they are
`quarterons'.\cite{uslongI}
 This can be seen from the Chern Simons
theory\cite{uslongI} or from the fact that two of them together make a
skyrmion which is a fermion.

\subsection{Kosterlitz-Thouless Phase Transition}
\label{subsec:KT}

It is well known that the presence of vortex topological defects can
disorder the XY phase of the ground state.  This will occur even
at zero temperature due to quantum fluctuations if the layer separation
exceeds a critical value $d^*$.   One  could also conveniently tune
through this transition in a single sample with fixed $d$ by varying a gate
voltage to induce layer imbalance which will renormalize the spin
stiffness.
We focus here however on the effect
of finite temperature and thermally induced vortices in samples
which are otherwise ordered at zero temperature (because $d<d^*$).

Integrating out the massive $m^z$ fluctuations and going to finite
temperature, we are led to a classical XY model of the form shown
in Eq.(\ref{eq:2dxy}).  Hartree-Fock estimates of the spin stiffness
at finite layer separation give  values in the range 0.1-0.5 K
for typical
experimental sample parameters.  The Kosterlitz-Thouless phase transition
is an unbinding of vortex defects in the XY model and occurs at a temperature
given approximately by the value of the bare spin stiffness.
These unbound vortices will cause the long wavelength effective
spin stiffness in Eq.(\ref{eq:2dxy}) to renormalize discontinuously to zero
at the Kosterlitz-Thouless critical point.

In the usual way, the scale-invariant classical action of Eq.(\ref{eq:2dxy})
leads to a logarithmic interaction among vortices.  A gas  consisting
of M merons will have an energy of the form
\begin{equation}
E = M E_{\rm core}
- \sum_{i<j}^M n_i n_j 2\pi \rho_s  \ln{\left(\frac{R}{R_{\rm core}}\right)}
+ \sum_{i<j}^M q_i q_j\frac{e^2}{4\epsilon R_{ij}}.
\label{eq:meron-interaction}
\end{equation}
Here $E_{\rm core}$ is the meron core energy,\cite{foot:simplicity2}
  $R_{\rm core}$ is the meron core size,
$R_{ij}$ is the separation between the $i$th and $j$th merons,
$n_i$ is the vortex charge (winding number, $\pm1$) of the $i$th meron.
The last term is new and is unique to the present problem.  It represents
the Coulomb interaction among the fractional charges bound to the merons.
$q_i=\pm1$ is the sign of the electrical charge ($\pm e/2$)
of the $i$th meron.

The origin of the logarithmic interaction in a superconducting or superfluid
film is the {\em kinetic} energy stored in the supercurrents circulating around
the vortices due to order parameter phase gradients.
Here the logarithmic interaction arises from the loss of
Coulomb exchange {\em potential} energy in the presence of phase gradients
produced by the vortices.

The Coulomb interaction in Eq.(\ref{eq:meron-interaction})
 falls off more rapidly than the log interaction
(it is effectively one order higher in derivatives than the log interaction)
and so is perturbatively irrelevant at the Kosterlitz-Thouless critical
point.   That is, the Kosterlitz-Thouless temperature may be shifted
somewhat but the phase transition itself is unaffected.  However in
the limit of strong Coulomb effects (due to small $\epsilon$ or large
vortex fugacity) the global phase diagram becomes extremely rich. Among
other things there is a phase transition to a chiral state with
non-zero order parameter $\langle n_i q_i\rangle$ in which
vortex charge and electrical charge are no longer independent.
The rich physics and the novel phase diagram of
this model has been elucidated in an interesting paper by
Tupitsyn, Wallin, and Rosengren.\cite{wallin}

If found experimentally, this KT transition would be the first
finite temperature phase transition in a quantum Hall system.
All other transitions between plateaus, etc. are zero temperature
transitions because the vortices there (Laughlin quasiparticles)
do not interact logarithmically and are unconfined by an analog of
the Anderson-Higgs mechanism.\cite{SMGchap10,zhang_cs,lz,ODLRO}
As noted above the characteristic energy scale for $\rho_s$ is
0.1-0.5 K.   This is not a problem from the point of view of
experimentally achievable temperatures.  However, in order to have
the needed U(1) symmetry it is essential (as we discuss in
Section~\ref{sec:tunneling})  that the tunneling amplitude
between the layers be much smaller than this scale.  Such samples
have not yet been constructed, but could be in principle because the
tunneling amplitude falls off exponentially with layer separation while
the Coulomb interactions which control the stiffness $\rho_{s}$ fall
off only as a power law.  Nevertheless it will not be easy since present
samples are already close to the critical separation $d^*/\ell$.  The
latter could perhaps be circumvented by going to lower density samples and
correspondingly lower magnetic fields.

The standard experimental signatures of the KT transition are zero
linear response dissipation below $T_{\rm KT}$, but zero critical current,
and a characteristic jump in the exponent associated with the
non-linear response
\begin{equation}
V \sim I^p,
\end{equation}
from $p=1$ above the transition to $p=3$ just below the
transition.\cite{KTexpts}  Dissipation is caused by phase slips
associated with the motion of unconfined vortices.  A transport current
exerts a magnus force whose direction depends on the sign of the vorticity
and which moves the vortices at right angles to the current.
Below the transition
the vortices are confined into (vorticity) neutral pairs by their
logarithmic attraction and no longer couple to the supercurrent.

In a superconductor or superfluid at finite temperature there is always
some sort of `normal' fluid present (thermally excited Bogoljubov
quasiparticles, for example) which can produce dissipation.  However
there are no electric fields to couple to these particles since the
supercurrent shorts them out.  Thus there is no dissipation.
Analogous normal fluid excitations exist
in the quantum Hall systems we are studying here.
If we define normal fluid as anything which is charged but has no net
vorticity then we see that a pair of opposite vorticity, but like-charged
merons constitutes normal fluid that can be thermally activated.
The validity of the meron pair picture has been confirmed by numerical
calculations.\cite{kyang-macd}
In the pseudospin
superfluid channel (i.e., opposite electrical currents in each layer)
this normal fluid will not couple to the supercurrent since it is vortex
neutral.  The linear response dissipation will therefore drop to zero
below $T_{\rm KT}$ and all the other usual signatures of the 
Kosterlitz-Thouless transition will be present.

Doing transport  experiments in the pseudospin channel requires
separately contacting the two layers and having extremely high tunneling
resistance between the layers without moving them so far apart that they
decouple.  It would be easier to do the experiment in the ordinary
charge channel in which both layers are contacted simultaneously and
current flows in the same direction in both layers.  Unfortunately in
this channel there is a net electric field induced by the $\nu=1$ quantized
Hall resistivity and the normal fluid will couple to this producing
thermally activated dissipation.\cite{uslongI}
  While we would expect to see the
dissipation decrease rapidly towards zero below $T_{\rm KT}$, it would not
show any discontinuity as it would in the pseudospin channel.  This is
the same mechanism that causes the small but non-zero
thermally activated dissipation in ordinary quantum Hall
plateaus.

The energy for an excitation of `normal fluid' consisting of
a pair of like-charged but opposite vorticity merons given by 
Eq.(\ref{eq:meron-interaction}) is minimized at a separation of
\begin{equation}
R_0 =  \frac{e^2}{8\pi\epsilon\rho_{\rm s}}.
\label{eq:meron_separation}
\end{equation}
For typical values of $\rho_{\rm s}$, $R_0 \ge 10 \ell$, and so is
large enough to justify the field-theoretic continuum approximations
used in deriving Eq.(\ref{eq:meron-interaction}).

\section{Tunneling Between the Layers}
\label{sec:tunneling}

A finite tunneling amplitude $t$ between the layers breaks the $U(1)$ symmetry
\begin{equation}
H_{\rm eff} = \int d^2r
\left[\frac{1}{2}\rho_s \vert\nabla\varphi\vert^2 - \frac{t}{2\pi\ell^2}
 \cos{\varphi}\right]
\label{eq:H_eff}
\end{equation}
by giving a preference to symmetric tunneling states.  This can be
seen from the tunneling Hamiltonian
\begin{equation}
H_{\rm T} = - t \int d^2r
\left\{\psi_\uparrow^\dagger ({\bf r}) \psi_\downarrow ({\bf r})
+ \psi_\downarrow^\dagger ({\bf r}) \psi_\uparrow ({\bf r})\right\},
\label{eq:tunnel}
\end{equation}
which can be written in the pseudospin representation as
\begin{equation}
H_{\rm T} = - 2t \int d^2r \,S^x({\bf r}).
\end{equation}
(Recall that the eigenstates of $S^x$ are symmetric and
antisymmetric combinations of up and down.)

We can shed some more light on the spontaneous symmetry breaking by
considering the tunneling Hamiltonian $H_{\rm T}$ in Eq.(\ref{eq:tunnel})
as a weak perturbation. Naively, since particle number is separately
conserved in each layer for $t=0$, one might expect
\begin{equation}
\lim_{t\longrightarrow 0} \frac{1}{t} \left\langle\psi\left|H_{\rm
T}\right|\psi\right\rangle = 0.
\end{equation}
That is, one might expect that the first-order term in the perturbation
series for the energy shift due to $t$ to vanish.  Instead however,
we find that
the energy shifts {\it linearly\/} in $t$
\begin{eqnarray}
\lim_{t\longrightarrow 0} \lim_{A\longrightarrow\infty}
\frac{1}{tA} \left\langle\psi\left|H_{\rm T}\right|\psi\right\rangle
&=& \lim_{t\longrightarrow 0} \lim_{A\longrightarrow\infty}
-\frac{1}{tA}
\left\langle\psi\left|\int d^2 r\, 2S^x({\bf r})\right|\psi\right\rangle\cr
&=& -m^x,
\end{eqnarray}
where $A$ is the system area, and $m^x$ is, by definition,
the magnetization which is
the system's order parameter.\cite{order-of-limits} If the interlayer
spacing $d$ is taken to be zero, one can readily show\cite{uslongI} that the
variational wave function in Eq.(\ref{eq:variational}) is exact,
hence $\lim_{t\longrightarrow 0} m^x = 1$, and $t = \Delta_{SAS}/2$.
For finite $d$,
Eq.({\ref{eq:variational}) is no longer exact and quantum
fluctuations will\cite{uslongI} reduce the magnitude of $m^x$ and
we must renormalize the hopping parameter $t$
appropriately.

As the layer separation $d$ increases, a critical point $d^*$ will be
reached at which the magnetization vanishes and the ordered phase is
destroyed by quantum fluctuations.\cite{macd-expt,usPRL,uslongI}
This is illustrated in Fig.[\ref{fig:qhe-noqhe}] and discussed in more detail
from the experimental point of view in the Chapter
by J. P. Eisenstein.\cite{JPEchap}
For {\it finite\/} tunneling $t$,
the collective mode becomes massive and quantum fluctuations will be
less severe.\cite{uslongII}
  Hence the phase boundary in Fig.[\ref{fig:qhe-noqhe}]
curves upward
with increasing $\Delta_{\rm SAS}$.
For $\Delta_{SAS}=0$
the destruction of long-range order and the charge excitation gap
are intimately related and occur simultaneously at $d^*$ and zero
temperature.  For finite $\Delta_{SAS}$ the system
always has non-zero $m^x$ even in the phase with zero charge gap.

We have already seen that finite layer separation reduces the pseudospin
symmetry from SU(2) to U(1).
The introduction of finite tunneling amplitude destroys the U(1) symmetry
and makes the simple vortex-pair configuration extremely expensive,
thereby destroying the KT transition.
To lower the energy, the system distorts the spin deviations into a domain
wall or `string' connecting the vortex cores as shown in
Fig.[\ref{fig:meron_string}].  The spins are oriented in the ${\bf\hat x}$
direction everywhere except in the shaded domain line region where they
tumble rapidly through $2\pi$.
The domain line has a fixed energy per unit length and so the vortices
are now confined by a linear `string tension' rather than
logarithmically.  We can estimate the string tension by examining the
energy of a domain line of infinite length.  The optimal form
for a domain line lying along the $y$ axis is given by
\begin{equation}
\varphi({\bf r}) = 2 \arcsin{[\tanh{(x/\xi)}]},
\end{equation}
where the characteristic width of the string is
\begin{equation}
\xi = \left[\frac{2\pi\ell^2\rho_s}{t}\right]^\frac{1}{2}.
\end{equation}
The resulting string tension is\cite{gruner}
\begin{equation}
T_0 = 8 \left[\frac{t\rho_s}{2\pi\ell^2}\right]^\frac{1}{2} =
\frac{8 \rho_s}{\xi}.
\label{eq:tension_0}
\end{equation}
Provided the string is long enough ($R \gg \xi$), the total energy of
a segment of length $R$ will be well-approximated by the expression
\begin{equation}
E_{\rm pair}' = 2E_{\rm mc}' + \frac{e^2}{4\epsilon R} + T_0R.
\label{string_pair}
\end{equation}
The prime on $E_{mc}$ in Eq.(\ref{string_pair}) indicates
that the meron core energy can
depend $\Delta_{SAS}$.  $E_{\rm pair}'$ is minimized
at $R=R_{0}' \equiv \sqrt{e^2/(4\epsilon T_0)}$.
Note that apart from the core energies, the charge gap at fixed
layer separation (and hence fixed $\rho_s$) is proportional to
$T_0^{1/2} \propto t^{1/4} \sim \Delta_{SAS}^{1/4}$
which contrasts with the case of free electrons, for which
the charge gap is proportional to $\Delta_{SAS}$.  Note that because
the exponent $1/4$ is so small, there is an extremely
rapid initial increase in the charge gap when tunneling first
becomes important.  (See below.)  
Note that the present classical analysis ignores the
stabilizing effect of the tunneling-induced gap on the
quantum fluctuations of the pseudospin magnetization.

The crossover between the meron-pair pseudospin texture
which holds for $t \equiv 0$ and the domain line string pseudospin
texture described above occurs at a finite value of $t$ which we
can estimate by the following argument.  
Recall from Eq.(\ref{eq:meron_separation}), that the equilibrium
separation of a meron pair in a system with no tunneling is defined to be
$R_0$.  For $R_{0}' > R_0$ the
vortices are already bound by the logarithmic attraction due to
the gradient energy before the linear attraction due to the hopping
becomes important at larger separations.  In this regime
tunneling does not play an important role in determining the nature
of the lowest energy charged pseudospin texture.  As $t$ increases
$R_{0}' \propto t^{-1/4}$ decreases and will eventually reach $R_0$
which is, of course,  independent of $t$.  Since
\begin{equation}
\frac{R_{0}'}{R_0} = 
\left(\frac{ 2 \pi^2 \rho_s}{e^2/(\epsilon  \xi)} \right)^{1/2}
= \frac{\pi \xi}{4 R_{0}'},
\label{eq:crossover}
\end{equation}
the characteristic width of the domain line becomes comparable to
$R_{0}'$ in the same range of $t$ values where $R_{0}'$ and $R_0$
become comparable.  We may conclude that the nature of the
charged pseudospin texture crosses over directly from the
meron pair form to the finite length domain line string form for
$\rho_s/[e^2/(\epsilon \xi)] 
\sim 1/25$, or equivalently for $t \sim t_{\rm cr}$ where
\begin{equation}
\frac{t_{\rm cr}}{(e^2/\epsilon\ell)} = 3.9 \times 
10^3 {\left[\frac{\rho_s}{e^2/(\epsilon\ell)}\right]}^{3}.
\label{eq:tcross}
\end{equation}
The crossover tunneling amplitude is thus typically smaller than
$ 5 \times 10^{-4} [e^2/(\epsilon \ell)]$.   Typical tunneling amplitudes in
double-layer systems are smaller than $ \sim 10^{-1} [e^2/(\epsilon\ell)]$
and can be made quite small by adjusting the barrier material or
or making the barrier wider.  Nevertheless, it seems likely that
$t$ will be larger than $t_{\rm cr}$ except for samples which are carefully
prepared to make $t$ as small as possible.  As $t$ increases beyond
$t_{\rm cr}$, $R_{0}'$ will continue to decrease.  When $R_{0}'$ becomes
comparable to the microscopic length, $\ell$, the description given
here will become invalid and the lowest energy charged excitations
will have single-particle character.  However, the domain-wall string
picture of the charged pseudospin texture has a very large
range of validity since $R_{0}' \propto t^{-1/4}$ decreases
very slowly with increasing $t$ at large $t$.
Writing $R_{0}' \sim [e^2 / (\epsilon 8\pi \rho_s)]
(t_{\rm cr}/t)^{1/4}$ we find that $R_{0}' \sim \ell$ only for
$ t \sim 10^{-2} [(e^2/\epsilon\ell)^2/\rho_s ]$.  
Using typical values of
$\rho_{\rm s}$ we see that the charged excitation crosses over to single
particle character only when the hopping energy $t$ becomes
comparable to the microscopic interaction energy scale.  The various
regimes for the charge excitations of double-layer systems are
summarized in Table~\ref{table:regimes}.  Almost
all typical double-layer systems lie within the regime of the
domain-wall-string pseudospin texture charge excitation, and hence
will not exhibit a true Kosterlitz-Thouless phase transition.

\section{Parallel Magnetic Field in Double Layer Systems}
\label{sec:parallelB}

Murphy et al.\cite{murphyetal} have shown that the charge gap in double layer
systems is remarkably sensitive to the application of relatively weak magnetic
fields $B_\parallel$, oriented in the plane of the 2D electron gas.
Experimentally this field component is generated by
slightly tilting the sample relative to the magnetic field orientation.
Tilting the field (or sample) has
traditionally been an effective method for identifying effects due to
(real) spins because orbital motion in a single-layer 2DEG system
is primarily\cite{nickila} sensitive to $B_\perp$, while
the (real) spin Zeeman splitting is proportional to the full magnitude of
$B$.  Adding a
parallel field component will tend to favor more strongly
spin-polarized states.
For the case of the double layer $\nu =1$ systems studied by
Murphy et al., \cite{murphyetal} the
ground state is known to already be an isotropic ferromagnetic state  of
the {\it true spins} and the addition of a parallel field would not,
at first glance,
be expected to influence the low energy states since they are already
fully spin-polarized. (At a fixed Landau level filling factor,
$B_\perp$ is fixed and so the total $B$
and the corresponding Zeeman energy increase with tilt).
Nevertheless experiments\cite{murphyetal} have shown that
these systems are very sensitive to $B_\parallel$.  The activation energy
drops rapidly (by factors varying from two up to an order-of-magnitude
in different samples) with increasing $B_\parallel$.
At $B_\parallel = B_\parallel^\ast$ there appears to be
a phase transition to a new state whose activation gap is approximately
independent of further increases in $B_\parallel$.

The effect of $B_\parallel$ on the {\it pseudospin\/} system
can be visualized in two different pictures, one microscopic, the other
macroscopic.  We will present the latter here.  The technical details
of the former are described elsewhere.\cite{usPRL,uslongII}
Recent work presents a discussion of higher Landau levels.\cite{jianyang}

We use a gauge in which ${\bf B}_\parallel =
{\bf\nabla}\times {\bf A}_\parallel$ where
${\bf A}_\parallel = B_\parallel (0,0,x)$.
In this gauge the vector potential points in the ${\bf\hat z}$ direction
(perpendicular to the layers) and varies with position $x$ as one
moves parallel to the layers.
In this gauge,
the only change in the Hamiltonian caused by the parallel field is
in the term which describes tunneling between layers.  As
an electron tunnels from one layer to the other it moves along the
direction in which the vector potential points and
so the tunneling matrix element acquires a position-dependent phase
$t\rightarrow t~e^{iQx}$ where $Q=2\pi /L_\parallel$ and $L_\parallel =
\Phi_0/B_\parallel d$ is the length associated with one flux quantum
$\Phi_0$ between the layers (defined in Fig.[\ref{fig:parallelB}]).
This modifies the tunneling Hamiltonian to $H_T=-\int
d^2r~{\bf h}({\bf r})\cdot {\bf S}({\bf r})$ where ${\bf h}({\bf r})$
`tumbles': {\it i.e.},
${\bf h}({\bf r})=2t~(\cos{Qx},\sin{Qx},0)$.  The effective XY
model now becomes
\begin{equation}
H=\int d^2r~\Biggl\{ \frac{1}{2}~\rho_s\vert {\bf\nabla}\varphi\vert^2 -
{t\over 2\pi\ell^2}~\cos{[\varphi ({\bf r}) - Qx]}\Biggr\} ,
\label{eq:140smg}
\end{equation}
which is precisely the Pokrovsky-Talapov (P-T) model\cite{bak} and has a
very
rich phase diagram.  For small $Q$ and/or small $\rho_s$ the phase obeys
(at low temperatures)
$\varphi
({\bf r})\equiv Qx$; the order parameter
 rotates commensurately with the pseudospin Zeeman
field.  However, as $B_\parallel$ is increased,
the local field tumbles too rapidly and a continuous
phase transition to an incommensurate state with broken
translation symmetry occurs.  This is because
at large $B_\parallel$, it costs too much exchange
energy to remain commensurate and the system rapidly gives up the tunneling
energy in order to return to a uniform state ${\bf\nabla}\varphi\approx 0$
which becomes independent of $B_\parallel$.  As explained in further detail
below we
find that the phase transition occurs at zero temperature
for\cite{usPRL,uslongII}
\begin{equation}
B_\parallel^\ast = B_\perp ~ (2 \ell / \pi d) ( 2 t / \pi \rho_s)^{1/2}.
\label{eq:critparfield}
\end{equation}
Using the parameters for the sample
of Murphy et al.\cite{murphyetal} with weakest 
tunneling\cite{changed_params} ($\Delta_{\rm SAS}=0.45$ K)
and neglecting quantum fluctuation
renormalizations of both $t$ and $\rho_s$ (i.e., using Hartree Fock)
 we find that the critical field for the transition is
$ \approx 1.3 {\rm T}$ which is slightly
larger than the observed value\cite{murphyetal,changed_params} of $0.8$ T,
but still correctly corresponds to a  very large length $L_\parallel$. 
A graphical comparison showing the qualitative agreement between
the predicted and observed values of the critical tilt angle for
several samples is
shown in Fig.(15) of the Chapter by Eisenstein in this volume.

In addition to the Hartree-Fock calculations described here,
we have made numerical exact diagonalization studies on small systems to
find the critical value of the parallel field.   Although it
is difficult to extrapolate the results to the thermodynamic limit, they 
do confirm that at finite layer separation,
quantum fluctuations can reduce the predicted 
critical parallel field.\cite{uslongII}

As previously mentioned,
the observed value $B_\parallel^\ast =0.8{\rm T}$
corresponds in
these samples to a large value for $L_\parallel$: $L_\parallel /\ell\sim
20$ indicating that the transition is highly collective in nature.
We emphasize again that these very large length scales are possible in
a magnetic field  only
because of the interlayer phase coherence in the system associated with
condensation of a {\it neutral\/} object.

Having argued for the existence of the commensurate-incommensurate
transition, we must now connect it to the experimentally observed
transport properties.
In the commensurate phase, the order parameter tumbles more and more rapidly
as $B_\parallel$ increases.  As we shall see below,
it is this tumbling which causes
the charge gap to drop rapidly.  In the incommensurate phase, the state
of the system is approximately independent of $B_\parallel$ and this
causes the charge excitation gap to saturate at a fixed value.

Recall that in the presence of tunneling, the cheapest charged excitation
was found to be a pair of vortices of opposite vorticity and like charge
(each having charge $\pm 1/2$) connected by a domain line
with a constant string tension.  In the absence of $B_\parallel$ the
energy is independent of the orientation of the string.  The effect of
$B_\parallel$ is most easily studied by changing variables to
\begin{equation}
\theta({\bf r}) \equiv \varphi({\bf r}) - Qx.
\end{equation}
This variable is a constant in the commensurate phase but not in
the incommensurate phase.  In terms of this new variable,
the P-T model energy is
\begin{equation}
H=\int d^2r~\Biggl\{ \frac{1}{2}~\rho_s [(\partial_x\theta + Q)^2
+ (\partial_y)^2]
- {t\over 2\pi\ell^2}~\cos{(\theta)}\Biggr\} .
\label{eq:140asmg}
\end{equation}
We see that $B_\parallel$ defines a preferred direction in the problem.
Domain walls will want to line up in the $y$ direction and contain
a phase slip of a preferred sign ($-2\pi$ for $Q>0$) in terms of
the field $\theta$.  Since the extra term induced by $Q$ represents a total
derivative, the optimal form of the soliton solution is unchanged.  However the
energy per unit length of the soliton,
which is the domain line string tension,
decreases linearly with $Q$ and hence $B_\parallel$:
\begin{equation}
T = T_0\left[1 - \frac{B_\parallel}{B_\parallel^*}\right],
\end{equation}
where $T_0$ is the tension in the absence of parallel B field
given by Eq.(\ref{eq:tension_0}),
and
$B_\parallel^*$ is the critical parallel field at which the string tension
goes to zero.\cite{comment_renorm}  We thus see that by tuning $B_\parallel$
one can conveniently control the `chemical potential' of the domain lines.
The domain lines condense and
the phase transition occurs (in mean field theory)
 when the string tension becomes negative.

Recall that the charge excitation gap is given by the energy of
a vortex pair separated by the optimal distance 
$R_0 = \sqrt{e^2/(4\epsilon T)}$.
\null  From Eq.(\ref{string_pair}) we have that the energy gap far on
the commensurate side of the phase transition is given by
\begin{eqnarray}
\Delta &=& 2 E_{\rm mc}' + \left[\frac{e^2T}{\epsilon} \right]^\frac{1}{2}\cr
&=& 2 E_{\rm mc}' + 
\sqrt{\frac{e^2T_0}{\epsilon}}\left[1-\left(\frac{B_\parallel}{B_\parallel^*}
\right)
\right]^\frac{1}{2}.
\end{eqnarray}
As $B_\parallel$ increases the reduced string tension allows the Coulomb
repulsion of the two vortices to stretch the string and lower the energy.
Far on the incommensurate side of the phase transition,
the possibility of interlayer tunneling becomes irrelevant.  From
 the discussion of the previous section one can argue that the ratio
of the charge gap at $B_{\parallel} = 0 $ to the charge
gap at $B_{\parallel} \to \infty$ should be given (very roughly) by
\begin{equation}
\frac{\Delta_0}{\Delta_{\infty}} = (t/t_{\rm cr})^{1/4} \approx
\frac{(e^2/\epsilon\ell)^{1/2} t^{1/4}}{ 8 {\rho_s}^{3/4}}.
\label{eq:gapratio}
\end{equation}
Putting in typical values of $t$ and $\rho_s$ gives gap ratios in the
range $\sim 1.5-7$ in qualitative
agreement with experiment.  According to the discussion of
the previous section, gap ratios as large as
$\sim (t_{\rm max}/t_{\rm cr})^{1/4}
\sim 0.07 (e^2/\epsilon\ell) / \rho_s$,
can be expected in the regime where the pseudospin texture picture applies.
Here $t_{\rm max}$ is the hopping parameter at which the crossover to
single-particle excitations occurs.  Thus gap ratios as large as an order
of magnitude are easily possible.  Of course, all the discussion here
neglects orbital effects (electric subband mixing)
within each of the electron gas layers, and
these will always become important at sufficiently strong parallel fields.
[Note also that for $d$ near $d^*$, the system will have enhanced
sensitivity to renormalization of parameters by electric subband mixing in
tilted fields.]

It should be emphasized that only this highly collective picture
involving large length scale distortions of topological defects can possibly
explain the extreme sensitivity of the charge gap to small tilts of the
B field.  Recall that at $B_\parallel^*$, the tumbling length $L_\parallel$
is much larger than the particle spacing and the magnetic length.  Simple
estimates of the cost to make a local
one-body type excitation (a spin-flip pair for example) shows that
the energy decrease due to $B_\parallel$ is extremely small since
$\ell/L_\parallel$ is so small.
Numerical exact-diagonalization calculations on small systems
confirm the existence of this
phase transition and show that the fermionic excitation gap
drops to a much smaller value in the incommensurate phase.\cite{uslongII}

The collective excitation modes of the system in the commensurate
and incommensurate phases appear to be quite
interesting.\cite{parallel_coll_modes}  As mentioned previously, the
tumbling of the order parameter may produce a self-grating effect which
will allow one to tune the wavevector which couples to optical probes.

All of our discussion of the phase transition in a parallel field
has been based on mean-field theory.  Close to the phase transition,
thermal fluctuations will be important.  At finite temperatures
there is no strict phase transition at $B_\parallel^\ast$ in the
the P-T model.  However there is a finite temperature KT phase transition
at a nearby $B_\parallel > B_\parallel^\ast$.  At finite temperatures
translation symmetry is restored\cite{bak} in the incommensurate phase
by means of dislocations in the domain string structure.
Thus there are two separate KT transitions in this system,
one for $t=0$, the other for $t\neq 0$ and $B_\parallel > B_\parallel^\ast$.
Recently Read\cite{read_2layer} has studied this model at finite
temperatures in some detail and has shown
that just at the critical value of $B_\parallel$ there should be
a square-root singularity in the charge gap.  The existing data does
not have the resolution to show this however.
M. P. A. Fisher has pointed out\cite{mpaf_comment}
that at zero-temperature the commensurate-incommensurate
phase transition must be treated quantum mechanically.
It is necessary to take account of the
world sheets traced out by the time evolution of the strings which
fluctuate into existence due to quantum zero-point motion.
He has also pointed out that the inevitable random variations
in the tunneling amplitude with position,
which we have not considered at all here, cause a relevant perturbation.


\section{Summary}
\label{sec:summary}

We have discussed the origin of spontaneous ordering in multicomponent
fractional quantum Hall effect systems.  For `real' spin at filling
factors $\nu=1$ this spontaneous ferromagnetism induces a very large
charge excitation gap even in the absence of a Zeeman gap.  The charged
excitations are interesting topological `skyrmion'-like objects.
Magnetic resonance experiments\cite{barrett_prl,barrett_science}
 have confirmed this remarkable
picture which was developed analytically by Sondhi et al.\cite{sondhi}
to explain the numerical results of Rezayi.\cite{rezayi}

We have discussed in detail a pseudospin analogy
which shows how spontaneous interlayer coherence in double
layer quantum Hall systems arises.  This coherent XY phase order occurring
over long length scales
is essential to explain the experimental
observations of Murphy et al.\cite{murphyetal} described in the Chapter
in this volume
 by J. P.  Eisenstein.  The essential physics is condensation
of a charge-neutral bosonic order parameter field (pseudospin
magnetization).  This condensation controls the charge excitation gap
and is very sensitive to interlayer tunneling and parallel magnetic field.

We summarize a portion of this rich set of phenomena
in the schematic zero-temperature phase diagram of double layer systems
shown in Fig.[\ref{fig:phase-diagram}]. First consider the plane
with $\Delta_{\rm SAS}=0$ (zero tunneling).
We have argued that the system develops spontaneous interlayer
phase coherence despite the fact that the tunneling amplitude is zero.
 However if the layer spacing $d$ exceeds a
critical value $d^*$, the system is unable to support a state with strong
interlayer correlations and the spontaneous $U(1)$ symmetry breaking is
destroyed by quantum fluctuations.\cite{macd-expt}  At this same point
we expect the fermionic gap $\Delta\mu$ to collapse.  Little is known about
the nature of this quantum transition which can be viewed as arising from
proliferation of quantum induced vortices (merons).  This is similar to the
quantum XY model, but in the present case, the merons are fractional-statistics
anyons which will presumably change the universality class of the
transition.\cite{uslongI}

At finite tunneling, the $U(1)$ symmetry is destroyed and the quantum
fluctuations are gapped and hence stabilized. This causes the critical
layer spacing to increase as shown in Fig.[\ref{fig:phase-diagram}].

The third axis in the figure is the tilt of the magnetic field.
Magnetic flux between the layers causes the order parameter to want to
`tumble'.  For small tilts, the system is in a commensurate phase with the
order parameter tumbling smoothly.  However above a critical value of the
parallel field, this tumbling costs too much exchange energy, and the
system
goes into an incommensurate phase which spontaneously breaks translation
symmetry (in the absence of disorder).  This phase transition has been
observed by Murphy et al.\cite{murphyetal,JPEchap} through the
rapid drop in the charge gap as the field is tilted.  We have presented
arguments that the charge gap is determined by the cost of creating a
highly collective object:  a pair of fractionally charged vortices
connected by a string.  It is the decrease of the string tension with tilt
which causes the extreme sensitivity to small tilts.

In addition to all this, there is (for zero tunneling) a finite temperature
Kosterlitz-Thouless phase transition.   If observed experimentally,
this would represent the first finite temperature phase
transition in a quantum Hall system.

We have tried to keep the present discussion as qualitative as possible.
The reader is directed to the many references for more detailed discussion
of technical points.  We close by noting that
there are still many open questions in this very rich field
concerning such things as edge states
in multicomponent systems, proper treatment of quantum fluctuations for
the highly collective excitations which appear to exist in these systems,
and the nature of the
phase transition at the critical value of the layer  separation.


\section{Acknowledgments}
\label{sec:acknowlegments}

The work described here is the result of an active and ongoing
collaboration with our colleagues L.~Brey, R.~C\^ot\'e,
H.~Fertig, K.~Moon, H.~Mori, Kun~Yang, Lotfi Belkhir,
L.~Zheng, D.~Yoshioka, and Shou-Cheng Zhang.
It is a pleasure to acknowledge numerous useful conversations with
D. Arovas, S. Barrett, G. Boebinger, J. Eisenstein, Z.F. Ezawa,
M. P. A.  Fisher, I. Iwazaki,
T.-L. Ho, J. Hu, D. Huse, D.-H. Lee, S. Q. Murphy, A. Pinczuk, M. Rasolt,
N. Read, S. Renn, M. Shayegan, S. Sondhi, M. Wallin, X.-G. Wen,
Y.-S. Wu, and A. Zee.
The work at Indiana University was supported by NSF DMR-9416906.



\addcontentsline{toc}{part}{Tables}

\begin{table}
\caption[]{Generalized Laughlin states for two component systems ($S$ is
the total spin quantum and $\ast$ denotes a state which is not an
eigenstate of ${\bf S}_T^2$).  The nominal filling factors $\nu_\uparrow$
and $\nu_\downarrow$ are shown in
parentheses for the ferromagnetic $\{m,m,m\}$ states because these are are not
unique (only their sum $\nu$ is fixed).
(After Ref.[\citen{macdsurface}]).}
\label{table1}
\begin{tabular}{lllllll}
$m$ & $m'$ & $n$ & $\nu_\uparrow$ & $\nu_\downarrow$ & $\nu$ & $S$\\
\tableline
 1 & 1 & 0 & 1    & 1    & 2    & 0\\
 1 & 1 & 1 & (1/2)  & (1/2)  & 1    & $N/2$\\
 1 & 3 & 0 & 1    & 1/3  & 4/3  & $N/4$\\
 1 & 5 & 0 & 1    & 1/5  & 6/5  & $N/3$\\
 3 & 3 & 0 & 1/3  & 1/3  & 2/3  & $\ast$\\
 3 & 3 & 1 & 1/4  & 1/4  & 1/2  & $\ast$\\
 3 & 3 & 2 & 1/5  & 1/5  & 2/5  & 0\\
 3 & 3 & 3 & (1/6)  & (1/6)  & 1/3  & $N/2$\\
 3 & 5 & 0 & 1/3  & 1/5  & 8/15 & $\ast$\\
 3 & 5 & 1 & 2/7  & 1/7  & 3/7  & $\ast$\\
 3 & 5 & 2 & 3/11 & 1/11 & 4/11 & $N/4$\\
 5 & 5 & 0 & 1/5  & 1/5  & 2/5  & $\ast$\\
 5 & 5 & 1 & 1/6  & 1/6  & 1/3  & $\ast$\\
 5 & 5 & 2 & 1/7  & 1/7  & 2/7  & $\ast$\\
 5 & 5 & 3 & 1/8  & 1/8  & 1/4  & $\ast$\\
 5 & 5 & 4 & 1/9  & 1/9  & 2/9  & 0\\
 5 & 5 & 5 & (1/10) & (1/10) & 1/5  & $N/2$\\
\end{tabular}
\end{table}

%

\begin{table}
\caption{Fractional charges for some two-component fractional
quantum Hall effect states.  $e^{X}_{X'}$ gives the contribution
to the charge from the $X'$ component in quasihole state$X$.}
\begin{tabular}{ccccccccc}
$m$ & $m'$ & $n$ & $e_{R}^{A}$ & $e_{G}^{A}$ & $e_{T}^{A}$ &
$e_{R}^{B}$ & $e_{G}^{B}$ & $e_{T}^{B}$\\ \hline
 1 & 1 & 0 & 1 & 0 & 1 & 0 & 1 & 1\\
 1 & 1 & 1 & ? & ? & 1 & ? & ? & 1\\
 1 & 3 & 0 & 1 & 0 & 1 & 0 & $1/3$ & $1/3$\\
 1 & 5 & 0 & 1 & 0 & 1 & 0 & $1/5$ & $1/5$\\
 3 & 3 & 0 & $1/3$ & 0 & $1/3$ & 0 & $1/3$ &
$1/3$\\
 3 & 3 & 1 & $3/8$ & $-1/8$ & $1/4$ &
$-1/8$ & $3/8$ & $1/4$\\
 3 & 3 & 2 & $3/5$ & $-2/5$ & $1/5$ &
$-2/5$ & $3/5$ & $1/5$\\
 3 & 3 & 3 & ? & ? & $1/3$ & ? & ? & $1/3$\\
 3 & 5 & 1 & $5/14$ & $-1/14$ & $2/7$ &
$-1/14$ & $3/14$ & $1/7$\\
 3 & 5 & 2 & $5/11$ & $-2/11$ & $3/11$ &
$-2/11$ &$3/11$ & $1/11$\\
 5 & 5 & 0 & $1/5$ & 0 & $1/5$ & 0 & $1/5$ &
$1/5$\\
 5 & 5 & 1 & $5/24$ & $-1/24$ & $1/6$ &
$-1/24$ & $5/24$ & $1/6$\\
 5 & 5 & 2 & $5/21$ & $-2/21$ & $1/7$ &
$-2/21$ & $5/21$ & $1/7$\\
 5 & 5 & 3 & $5/16$ & $-3/16$ & $1/8$ &
$-3/16$ & $5/16$ & $1/8$\\
 5 & 5 & 4 & $5/9$ & $-4/9$ & $1/9$ &
$-4/9$ & $5/9$ & $1/9$\\
 5 & 5 & 5 & ? & ? & $1/10$ & ? & ? & $1/10$\\
\end{tabular}
\label{table:fc}
\end{table}

\begin{table}
\caption[]{Charged Spin Texture Energies at $\nu_{T}=1$ for Double Layers
Systems with Tunneling. $\tilde{\rho}_{s}\equiv \rho_{s}/(e^{2}/\ell )$
and $\tilde{t}\equiv t/(e^{2}/\ell )$ where
$\rho_{s}$ is the pseudospin stiffness, $t$ is the renormalized tunneling
amplitude, $\ell$ is the magnetic length, $T_{0}=8\rho_{s}/\xi$ is
the soliton string tension and $\xi
=\left(\frac{2\pi\ell^{2}\rho_{s}}{t}\right)^{1/2}$ is the domain wall
width.}
\begin{tabular}{llll}
Regime & $\tilde{t} \leq 4 \times 10^{3}\tilde{\rho}_{s}^{3}$ & $4 \times
10^{3}\tilde{\rho}_{s}^{3} \leq \tilde{t} \leq 10^{-2}/\tilde{\rho}_{s}$
& $10^{-2}/\tilde{\rho}_{s} \leq t$\\ \tableline
Nature of Charged & Meron Pairs & Finite Length & Single
Particle\\
Excitations & & Domain Line Strings & Excitation\\ \tableline
Excitation Size & $\sim \frac{e^{2}}{8\pi\rho_{s}}$ & $\sim
\sqrt{\frac{e^{2}}{4T_{0}}} \propto t^{-1/4}$ & $\ell$\\ \tableline
Excitation Energy & $\sim 2\pi\rho_{s}$ & $\sim \sqrt{e^{2}T_{0}} \propto
t^{1/4}$ & $t$\\
\end{tabular}
\label{table:regimes}
\end{table}

\addcontentsline{toc}{part}{Figure Captions}

\begin{figure}
\caption[]{Collective mode dispersion
for a double-layer system at $\nu_T=1/2$
and $d/\ell=1.5$.  The energies of the inter-Landau-level modes are
measured from $\hbar \omega_c$.  The ground state is approximated by
the $(3,3,1)$ Halperin state.  The plotting symbols refer to the
following modes: triangles ($n=1$ sum mode); circle
($n=1$ difference mode); square ($n=0$ sum mode);
; diamond ($n=0$ difference mode).}
\label{fig:cm1}
\end{figure}

\begin{figure}
\caption[]{(a) Illustration of a simple spin-flip excitation which
creates a widely separated particle hole pair.
{(b)} A skyrmion spin configuration (shown in cross section).  The
spins gradually and smoothly rotate from up at the perimeter to down at the
origin in a circularly symmetric spin textural defect.  For the case of
Coulomb interactions, this object costs only 1/2 the energy of the simple
spin flip.}
\label{fig:spin-flip}
\end{figure}

\begin{figure}
\caption[]{Knight shift measurements of the electron spin polarization
of a 2DEG in the vicinity of filling factor $\nu=1$.
(After Barrett et al., Ref.[\citen{barrett_prl}]).}
\label{fig:knight-shift}
\end{figure}

\begin{figure}
\caption[]{Illustration of the path $\omega$ of a spin $S{\bf m}$
on a unit sphere.  When viewed from the origin of spin space, the path
subtends a solid angle $\Omega$ and so the path contributes a Berry's
phase $S\Omega$.}
\label{fig:berry}
\end{figure}

\begin{figure}
\caption[]{A smooth spin texture.  An electron moving along the boundary of
the region $\Gamma$ in real space
has its spin follow the path in spin space
along the boundary of the
region labeled $\omega$ in Fig.[\ref{fig:berry}].}
\label{fig:texture}
\end{figure}

\begin{figure}
\caption[]{Infinitesimal circuit in spin space associated with an
infinitesimal circuit in real space.}
\label{fig:stokes}
\end{figure}

\begin{figure}
\caption[]{Schematic conduction band edge profile for a
double-layer two-dimensional electron gas system.  Typical widths
and separations are $W\sim d\sim 100\hbox{\AA}$
and are comparable to the spacing
between electrons within each inversion layer.}
\label{fig:double-well}
\end{figure}

\begin{figure}
\caption[]{Phase diagram for the double layer QHE system (after
Murphy et al.\cite{murphyetal}).  Only samples whose parameters
lie below the dashed line exhibit
a quantized Hall plateau and excitation gap.}
\label{fig:qhe-noqhe}
\end{figure}

\begin{figure}
\caption[]{A process in double-layer two-dimensional electron
gas systems which encloses flux from the parallel component
of the magnetic field.  The quantum amplitude for such paths
is sensitive to the parallel component of the field.}
\label{fig:parallelB}
\end{figure}

\begin{figure}
\caption[]{The four flavors of merons.  These are vortices which are
right or left handed and have topological charge $\pm 1/2$.}
\label{fig:meron}
\end{figure}

\begin{figure}
\caption[]{Meron pair connected by a domain wall.  Each meron carries
a charge $e/2$ which tries to repel the other one.}
\label{fig:meron_string}
\end{figure}

\begin{figure}
\caption[]{Schematic zero-temperature phase diagram (with $d/\ell$
increasing downwards).  The lower surface is
$d^*$ below which $d > d^*$ and  the interlayer correlations are too weak
to support a fermionic gap, $\Delta\mu$.  The upper surface gives
$B_\parallel^*$, the commensurate-incommensurate phase boundary. As $d$
approaches $d^*$, quantum fluctuations soften the spin stiffness and
therefore increase
$B_\parallel^*$.}
\label{fig:phase-diagram}
\end{figure}


\begin{references}
\bibitem{realize:up-down} It is important to realize that if $\Phi [Z]$ is
antisymmetric among spin-up electrons and among spin-down electrons, the
expectation value of an operator which does not flip spins, can be
calculated as if the ``up" and ``down" spin electrons were
distinguishable thus avoiding all the complications associated with the
antisymmetrization.
\bibitem{foot:simplicity1} For simplicity
we write the wave function for the essentially equivalent
$\nu=1/2$ case rather than dealing with the complications of the
higher Landau level wave functions needed to describe $\nu=5/2$.
\bibitem{caveat1} The SMA is not reliable for compressible states or
for hierarchical incompressible states.
\bibitem{exceptnu1} An exception occurs for the $\Psi_{1,1,1}$
wavefunction for which $\nu =1$.  In this case there are no states
in the lowest Landau level to which $\Psi_{1,1,1}$ is coupled by
the total charge density operator.  The reasons for this will
become clearer later in this chapter.
\bibitem{rm} The correlation functions were obtained by using a 
hypernetted chain approximation as described in Refs.[\citen{rasolt}].
\bibitem{caveat2} The absorption strength of this mode
will be reduced by a factor of $\sim
(d/\cos(\theta) \lambda)^2$ where $\lambda$ is the wavelength of the
infrared light and $\theta$ is the angle between the propagation
direction and the normal to the 2D layers.
\bibitem{foot:hand} On
the other hand, we remind the reader that because of analyticity and
Pauli principle constraints on the Landau level wave functions,
Hund's rule does not always apply.  We saw for instance
in Section~\ref{sec:multi-psi} that in some cases, particularly those
with even denominator filling factors such as
$\nu=5/2$, the interaction energy can be
 optimized by a spin liquid singlet(-like) state.
\bibitem{precise_def}  The precise definitions of the skyrmion and
antiskyrmion energies in the presence of the long range Coulomb interaction
is subtle.  For a more complete discussion see Ref.[\citen{uslongI}].
\bibitem{foot:adiabatic} We require $\sigma_{xx}=0$ so that the flux
can be added adiabatically and reversibly.
\bibitem{nuhalf} A single-layer system at Landau level filling
factor $\nu=1/2$ has no charge gap but does show interesting
anomolies which may indicate that it forms a liquid
of composite fermions.  For a discussion of recent work see
the chapters in this volume by B. I. Halperin and by J. K. Jain.
\bibitem{sondhi_phrase}  This phrase was the original title of
Ref.[\citen{sondhi}] prior to the editorial process.
\bibitem{foot:simplicity2} For simplicity we
ignore the fact that different flavors may have different core energies.
\bibitem{order-of-limits}  Note the order of limits here.  The fact that
the order matters really defines what we mean by spontaneously broken
symmetry.
\bibitem{comment_renorm}  We are considering for the moment the case of zero
temperature so that $B_\parallel$ is unrenormalized by thermal fluctuations.
We also treat the problem classically.
\end{references}
\end{document}